\documentclass[twocolumn,showpacs,reprint,amsmath,amssymb,aps,prc,10pt]{revtex4-1}
\usepackage{graphicx}% Include figure files
\usepackage{dcolumn}% Align table columns on decimal point
\usepackage{bm}% bold math
\begin{document}
\title{Fragmentation of $^{14}$N, $^{16}$O, $^{20}$Ne, and $^{24}$Mg Nuclei 
at 290 to 1000 MeV/nucleon}% Force line breaks with \\

\author{C. Zeitlin}
\email{zeitlin@boulder.swri.edu}
\affiliation{Southwest Research Institute, Boulder, CO 80302}
\author{J. Miller}
\affiliation{Lawrence Berkeley National Laboratory, 1 Cyclotron Road, Berkeley, CA 94720}
\author{S. Guetersloh}
\affiliation{Department of Nuclear Engineering, Texas A{\&}M University, College Station, TX 77843}
\author{L. Heilbronn}
\affiliation{Department of Nuclear Engineering, University of Tennessee, Knoxville, TN 37996}

\author{A. Fukumura}
\affiliation{National Institute of Radiological Sciences, Chiba, Japan}%
\author{Y. Iwata}
\affiliation{National Institute of Radiological Sciences, Chiba, Japan}%
\author{T. Murakami}
\affiliation{National Institute of Radiological Sciences, Chiba, Japan}%

\author{S. Blattnig}
\affiliation{NASA Langley Research Center, Hampton, VA 23681}
\author{R. Norman}
\affiliation{NASA Langley Research Center, Hampton, VA 23681}

\author{S. Mashnik}
\affiliation{Los Alamos National Laboratory, Los Alamos, NM 87545}

\date{\today}% It is always \today, today,
\begin{abstract}
We report fragmentation cross sections measured at 0$^\circ$ for beams of $^{14}$N, $^{16}$O, 
$^{20}$Ne, and $^{24}$Mg ions, at energies ranging from 290 MeV/nucleon to 1000 MeV/nucleon.
Beams were incident on targets of C, CH$_{2}$, Al, Cu, Sn, and Pb, with the C and
CH$_2$ target data used to obtain hydrogen-target cross sections. Using methods
established in earlier work, cross sections obtained with both large-acceptance
and small-acceptance detectors are extracted from the data and when necessary
corrected for acceptance effects. The large-acceptance data yield cross sections 
for fragments with charges approximately half of the beam charge and above, 
with minimal corrections. Cross sections for lighter fragments are obtained
from small-acceptance spectra, with more significant, model-dependent corrections 
that account for the fragment angular distributions. Results for both charge-changing
and fragment production cross sections are compared to the predictions of the Los Alamos 
version of the Quark Gluon String Model (LAQGSM) as well as the NUCFRG2 and PHITS models. 
For all beams and targets, cross sections for fragments as light as He are compared to the models.
Estimates of multiplicity-weighted helium production cross sections are obtained from
the data and compared to PHITS and LAQGSM predictions. Summary statistics show that the 
level of agreement between data and predictions is slightly better for PHITS than for 
either NUCFRG2 or LAQGSM. 
\end{abstract}

\pacs{25.75.-q, 25.70.Mn, 25.60.Dz, 24.10.Lx, 98.70.Sa}% PACS, the Physics and Astronomy
\maketitle

\section{Introduction}

The Galactic Cosmic Rays (GCR) contain a small percentage of heavy ions that contribute substantially to the 
dose and dose equivalent received in spaceflight \cite{ncrp2006}, particularly in deep space outside 
the geomagnetosphere. As NASA's future plans are likely to include extended human missions in deep space, 
these exposures take on increased importance. A detailed understanding of the transport of 
these ions through matter is needed, as crew will typically be inside a modestly-shielded spacecraft, 
in habitats (conceivably with relatively thick shielding), or, in the case of a Mars mission, shielded 
by a combination of the CO$_2$ atmosphere and a habitat. Fragmentation cross sections play a key role 
in transport calculations and the resulting estimates of dose and dose equivalent behind shielding. 
These estimates can be highly uncertain \cite{Wilson2001} due to propagation of cross section uncertainties. 
This can, in turn, be a central factor in limiting mission duration or rendering certain mission scenarios 
unfeasible. It is therefore essential that an accurate and precise database of the nuclear interaction 
cross sections is available to modelers for both code development and validation purposes.

Historically, the space radiation community has focused on heavy ions such as iron. This is understandable, 
since the heavier GCR ions - iron in particular - contribute substantially to the dose and dose equivalent 
in unshielded deep space. However, from the standpoint of model completeness and reliability, it is also 
important that the fragmentation of lighter ions be well understood, since B, C, N, O, Ne, and Mg ions are
much more abundant than the heavier ions. Furthermore, beams of C \cite{HIMAC, GSI} and Ne ions \cite{Castro} 
have also been used in cancer therapy, where fragmentation plays a key role in limiting the (desired) localization 
of the dose to the tumor volume. We present cross sections for beams of $^{14}$N ions at 290 and 400 MeV/nucleon;
$^{16}$O at 290, 400, 600, and 1000 MeV/nucleon; $^{20}$Ne at 290, 400, and 600 MeV/nucleon; and $^{24}$Mg at 
400 MeV/nucleon. Charge-changing (sometimes referred to as ``total'' charge-changing) cross sections and 
fragment production cross sections have been extracted from energy-loss spectra measured with silicon detectors. 
The fragment production cross sections are partial cross sections. In 0$^\circ$ experiments such as this, each
event has a ``leading'' (highest-charged) fragment. Absent angular acceptance corrections, the sum of these
leading-fragment cross sections equals the charge-changing cross section for any given combination of beam ion 
and target.

As in our previously-reported fragmentation cross section data 
\cite{Zeitlin1997,Zeitlin2001,Zeitlin2007a,Zeitlin2007b,Zeitlin2008}, 
the charge-changing cross sections in the 250 to 1000 MeV/nucleon range are reproduced with reasonable accuracy
by geometric models that treat the nuclei as overlapping spheres. In some cases, slight energy dependence is observed in the data, 
and most of the models attempt to account for this, but on the whole it is a small effect and energy-independent models 
may suffice. In contrast, model predictions of the fragment production cross sections are, in general, not nearly as 
accurate as those for charge-changing cross sections. Here too there are subtle dependences on beam energy, and these 
are typically not well reproduced by the models. Older models (e.g., Nilsen et al. \cite{Nilsen1995}, NUCFRG2 
\cite{Wilson1994}, and EPAX2 \cite{epax2000}) approximate fragment cross sections as monotonically 
decreasing functions of the charge change $\Delta Z$ ($= Z_{\text{beam}} - Z_{\text{frag}}$), where $Z$ is the nuclear charge. 
In many instances, this is 
fairly accurate in an average sense, over a limited range of $\Delta Z$, but this approach cannot hope to reproduce 
important details seen in the data. Features missing in these older models include the enhanced production of even-$Z$ 
fragment species relative to odd-$Z$ species, suppression of F ($Z$ $=$ 9) production, and increases in cross sections for large 
$\Delta Z$'s. (For an example of these discrepancies, see Figure 15 of Ref. \cite{Zeitlin2008}.)

Other high-quality data in the literature (e.g., \cite{Webber1990b, Knott1996, Iancu2005}) report only
fragment cross sections for (approximately) $Z_{\text{beam}}$/2. In contrast, in this and other articles we report
the small charge change cross sections and also take the extra step to extract light-fragment production 
cross sections. This is achieved using spectra obtained with detectors placed far from the target (therefore 
subtending small acceptance angles). Acceptance corrections are made using a calculation that combines Goldhaber's 
formulation of fragment angular distributions \cite{goldhaber} with well-known Coulomb multiple scattering 
distributions, which are only important when high-$Z$ targets are used (Sn, Pb). The acceptance correction method 
has been shown to work well in our earlier published data \cite{Zeitlin2007a, Zeitlin2007b, Zeitlin2008}, based 
on the fact that cross sections for heavier fragments measured at large acceptance can be accurately reproduced by 
the corrected small-acceptance results. The main uncertainty in the calculation is associated with $\sigma_{0}$, 
the parameter in the Goldhaber model that controls the widths of the momentum distributions of outgoing fragments 
after the collision. The light-fragment cross sections allow model tests at large $\Delta Z$ (i.e., smaller impact 
parameters), where few previous comparisons have been made. This is a more stringent test of models than is possible 
using data dominated by peripheral collisions (leading to small charge changes). However, the limitations of the
experimental method, which does not account for non-leading light fragments, and the meaning of the reported 
cross sections are not entirely obvious and must be considered when making comparisons to models. These points 
will be further elucidated below.

There are some instances of overlap between the measurements presented here and those made by Webber 
et al. \cite{Webber1990a, Webber1990b}, and we present comparisons where data sets are sufficiently similar. 
Webber et al. obtained data on hydrogen and carbon targets, while our chosen targets span the periodic chart 
to allow for study of target mass dependences in the cross sections. 

The data presented here are drawn from a series of fragmentation experiments performed between 1995 and 2006.
Analysis of these data continues with the specific goal
of extracting light fragment cross sections. A parallel effort is in progress to validate and verify the nuclear 
physics models used in space radiation shielding applications \cite{Norman2008}. The progress of model development 
over time is being tracked by placing the models under configuration control, with automated validation benchmarks 
to enable comparisons as models are improved. Validation metrics are focused on the specific applications of interest,
and have been developed to enable comparisons of fragmentation models to the relatively sparse experimental 
database. The data shown here improve and extend that database.

\section{Experiments}

The cross sections have been obtained from several separate experiments which shared a common design. 
The $^{16}$O data at 600 and 1000 MeV/nucleon were obtained at the NASA Space Radiation Laboratory (NSRL) 
at the Brookhaven National Laboratory. All other experiments were performed at the Heavy Ion Medical 
Accelerator in Chiba (HIMAC) at the Japanese National Institute of Radiological Sciences. In all cases, 
we identify particles using deposited energy ($\Delta E$) signals from small-area silicon detectors centered 
on and normal to the beam axis. Detectors are positioned just upstream of the target position so that event 
samples can be limited to those with one and only one well-identified primary beam ion present incident
on the target, with a position close to the nominal beam axis. Other detectors are placed downstream of 
the target, at various distances with respect to the target-center position so that they subtend different 
acceptance angles and measure different spectra. The large acceptance detectors, typically placed so as to 
subtend forward cones with half-angles between 5$^{\circ}$ and 10$^{\circ}$, have 100\% acceptance for 
surviving primaries and all fragments whose species can be identified, which generally extends as far as 
charges $Z_{\text{frag}} \ge Z_{\text{primary}}/2$. For lower $\Delta E$'s, there appears to be poor resolution, but this 
is in fact caused by the many possible combinations of light fragments, which results in overlapping 
$\Delta E$ distributions. Detectors placed downstream so as to have small acceptance, on the order of 
1$^\circ$ to 2$^\circ$, are hit by many fewer fragments, and produce spectra in which all fragment 
species can be resolved. Given that the detectors are unsegmented, even at small acceptance there is 
some unavoidable ambiguity in interpretation of some of the fragment peaks, since certain combinations 
of light fragments can be indistinguishable from a single fragment of a heavier species (e.g., two He 
fragments with a particular velocity, detected in coincidence, produce a signal in the detector very 
close in amplitude to that of a single Li fragment at the same velocity).

Figure 1 is a schematic drawing of the arrangement of detectors on the beamline for the 400 MeV/nucleon 
$^{14}$N experiment. It is representative of the configurations used for all experiments. A detailed discussion 
of the experimental setup can be found in Zeitlin et al. \cite{Zeitlin2007a}. The acceptance 
angles, defined as the half-angle of the forward cone extending from the target center to the radius of 
the detector, are as indicated in the figure. The detectors were arranged in pairs to facilitate the 
data analysis, which depends on correlations between neighboring detectors. 

\begin{figure*}
\includegraphics[width=5.0in]{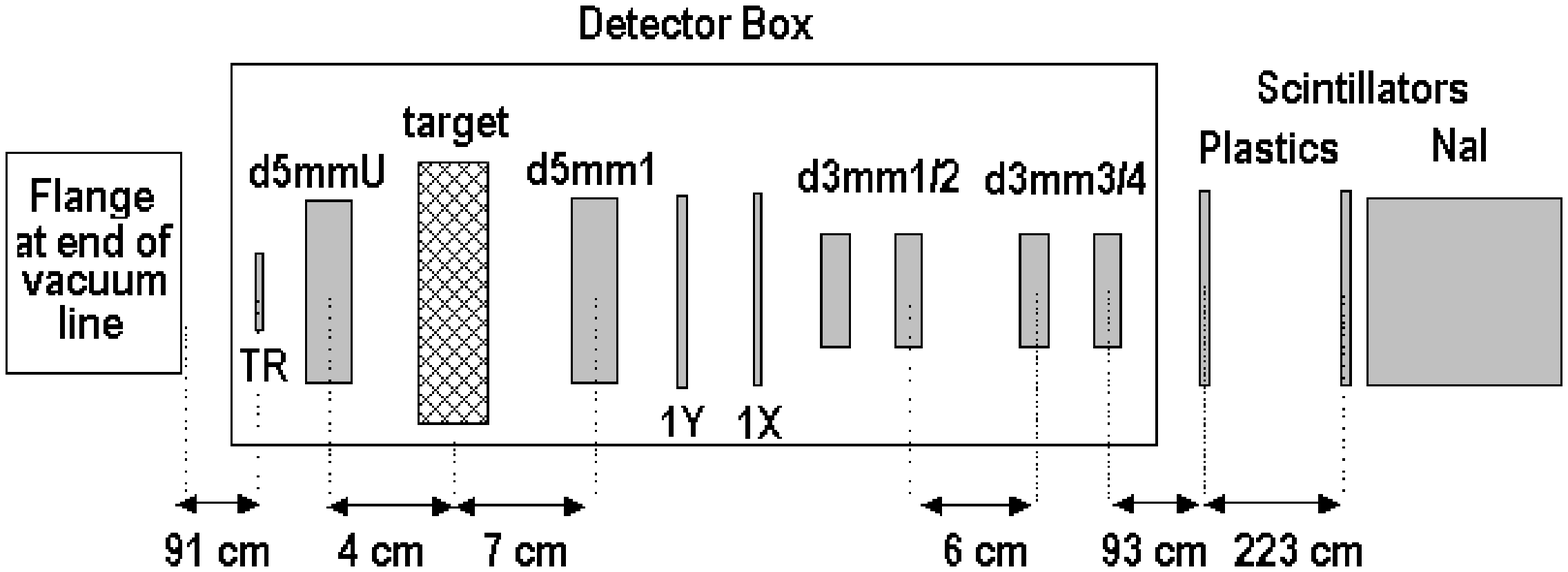}
\caption{\label{fig:fig1}Schematic diagram of the beamline configuration for the 400 MeV/nucleon $^{14}$N 
experiment. Spacing between detectors was 2 cm unless otherwise noted. Configurations for other
experiments were similar.}
\end{figure*}

\section{Event Selection and Particle Identification}

With the beam energies and targets used in these experiments, primary ions that survive traversal
of the target and projectile fragments generally have velocities that fall in a narrow range. 
This results in $\Delta E$ spectra with peaks that are well separated from one another, with $\Delta E$ 
$\propto$ $Z^2$. Both the number of visible fragment peaks in a particular spectrum, and the correspondence 
of the integrated counts in those peaks to cross sections, depend on the several factors described below.

As in earlier work (see references \cite{Zeitlin1997} - \cite{Zeitlin2008}), the CERN library program PAW \cite{Brun} 
was used to analyze the data. For every run, an initial event sample was selected by requiring that one and only one 
primary beam ion was seen in the detectors upstream of the target. For each detector pair downstream of the target, 
a scatter plot of $\Delta E$'s was made, and a cut contour (in some cases, multiple contours) was drawn to select those 
events having correlated pulse heights. These cuts remove events in which fragmentation occurred within either detector, 
along with events in which there was only partial charge collection in one of the detectors. For analysis of small-acceptance 
detector data, we selected those events in which the charge of the most forward-produced fragment could not be determined 
by the large-acceptance detectors, plus events corresponding to the two or three lightest fragment species that can be 
resolved at large acceptance. This provides overlap in the ranges of $Z$ measured in the two acceptances, which in turn 
allows us in a subsequent step to tune the acceptance model so that the cross sections match as closely as possible 
for the fragment species measured at both acceptances.

\subsection{Analysis Using Large-Acceptance Detectors}

There is a degree of subjectivity involved in drawing the cut contours in the scatter plots. The contour 
that defines the surviving primary ions is particularly important, since it directly affects the charge-changing 
cross section $\sigma_{cc}$. Since the fragment production cross sections are effectively normalized to 
$\sigma_{cc}$, this cut strongly affects all measurements for a given data set. The contours define the event 
selection efficiencies, that is, $N_{pass}(Z) = N_{true}(Z) \varepsilon (Z)$, where $N_{pass}(Z)$ is the number 
of events of a particular charge $Z$ within the contour, and $N_{true}(Z)$ is the ``real'' number of such particles. 
Our goal is to draw the contours so that all values of $\varepsilon (Z)$ are equal. However, there is no way to 
do this with perfect precision, and as a result the drawing of the contours is a source of systematic error that 
must be accounted for.

\subsection{Analysis Using Small-Acceptance Detectors}

For each run, we chose a subset of the events analyzed in the large-acceptance data for further analysis
using the small acceptance detectors. The subsamples consist mostly of events in which the $\Delta E$ in 
the large-acceptance detectors is in the unresolved portion of the spectrum. The remainder of the events chosen 
are those in which the charge as determined by the large-acceptance detectors is at the low end of what can be resolved. 
Events with well-correlated signals in the downstream detector pair are selected by drawing another cut contour in the
appropriate scatter plot. 

A typical large-acceptance fragment charge spectrum is shown in Figure 2a for the 400 MeV/nucleon $^{14}$N data,
with a small acceptance charge spectrum shown in Figure 2b. The comparison illuminates some basic physics. 
In the large-acceptance data, only two fragment species, B and C, can be resolved at large acceptance. In contrast, 
at small acceptance, considerable structure is visible in the charge histogram. The scale in this plot was determined
simply by scaling the square root of deposited energy so that the peak for C fragments (not shown) is centered at a 
charge of 6.0. As in previously-reported data sets, non-integer peaks are seen in addition to peaks for charges 1 
through 5. There is a peak near 4.5 due to detection of Be and He fragments in coincidence. There is also a broad peak 
centered near 3.5, due to the detection of three helium nuclei in coincidence. Since the beam ion was $^{14}$N, charge 
conservation allows for a projectile fragment of charge 1 ($^1$H or $^2$H) to also be present in these events. 
(The presence or absence of a charge 1 fragment contributes to the width of the $Z$ $\approx$ 3.5 peak.) The peak in the 
vicinity of charge 3 appears to be split, with a relatively sharp peak centered very close to 3.0 and a less-defined 
peak near 2.8. The former is likely due to Li fragments, and the latter to pairs of He fragments detected in coincidence,
some of which come from decay of any $^8$Be fragments. In most of our other data sets, the two peaks near charge 3
are not resolved, and though the separation here is not large, there do seem to be two distinct peak regions. Fitting 
two Gaussian distributions to this region yields a $\chi^2$ of 12.1 for 10 degrees of freedom, with one peak found at 
$Z$ = 2.76 $\pm$ 0.013 and the other at 2.96 $\pm$ 0.013 \footnote{With perfect scaling of $\Delta E$ to charge, we would 
expect these peaks to occur at $Z$ = 2.83 and 3.00, respectively. Both are shifted to slightly lower values because, on 
average, these lighter fragments have higher velocity when they exit the target than do the C fragments that are used 
to set the charge scale.}. Between charge 1.8 and about 2.6, there are three statistically-significant peaks at 1.94, 
2.15, and 2.46, likely corresponding to He alone, and He in coincidence with one and two H fragments, respectively. 
In the vicinity of charge 1, peaks appear at 0.88 and 1.36, corresponding to one and two singly-charged 
relativistic particles. Another notable feature of this spectrum is the suppression of charge 4 fragments. 
As mentioned above, any $^8$Be fragments created in the collisions instantaneously decay to two $^4$He ions, 
which may then be detected in the $Z$ $\approx$ 3.5 or $Z$ $\approx$ 2.8 peaks, both of which are far more populated 
than the charge 4 region, which extends from about 3.8 to 4.7. 

Additional complications arising in the interpretation of small-acceptance spectra are discussed elsewhere
\cite{Zeitlin2010}. It was noted in that article that, at least for one simulated data set (650 MeV/nucleon
$^{40}$Ar), some 80\% of He fragments were produced in association with heavier fragments. More generally, 
we can say that significant percentages of the lightest fragments are produced as non-leading fragments, and 
are difficult to account for. We note here that the light-fragment results presented below are
quite sensitive to the acceptance angles in the different experiments, and that large correction factors
are applied. The net result is that relative errors on the cross sections for the lightest fragments are 
large compared to those obtained for heavier fragments.

\begin{figure}
\includegraphics[width=3.37in]{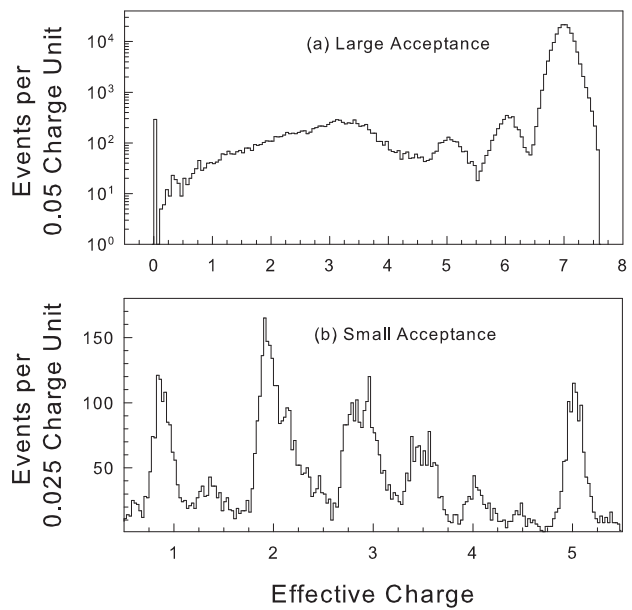}
\caption{\label{fig:fig2}Charge spectra at large (upper, Fig. 2a) and small (lower, 2b) acceptances,
for 400 MeV/nucleon $^{14}$N beam on a carbon target.}
\end{figure}

With the $^{14}$N beam, interpretation of the multiple-fragment peaks is relatively straightforward. Charge conservation 
dictates that (unlike in data sets with heavier beams) the charge 4 peak cannot be due to the detection of four He fragments 
in coincidence. When counting events by leading fragment, as is done in the cross section analysis, the events populating 
the peak near 3.5 are counted as He, and the only ambiguity is in assigning a charge to the events in the 2.6-3.5 region, 
as they may be either Li fragments or pairs of He fragments. For simplicity, we group them together in the cross section 
tables, but separately we will show estimates of the He:Li ratios.

\subsection{Experiment Acceptances}
The detector configuration changed between experiments, which were conducted over a period of several years
at different accelerator facilities. Table I shows the detector acceptance angles for the data sets presented
here. The acceptance angles are the half-angles of the forward cones as measured from the exact center of
the target.

\begin{table*}
\caption{\label{table:tab1}Acceptance angles for all data sets.}
\begin{ruledtabular}
\begin{tabular}
%{|p{55pt}|p{58pt}|p{64pt}|p{64pt}|p{70pt}|p{70pt}|p{70pt}|}
{c c c c}
Beam Ion & Energy at Extraction & Large Acceptance Angle & Small Acceptance Angle(s)\\
         & (MeV/nucleon)        &  (degrees)             & (degrees) \\
\hline
$^{14}$N & 290       &  5.7$^\circ$ & 1.7$^\circ$ \\
$^{14}$N & 400       &  9.8$^\circ$ & 3.8$^\circ$ \\
$^{16}$O & 290       &  5.7$^\circ$ & 1.7$^\circ$ \\
$^{16}$O & 400       &  6.7$^\circ$ & 2.5$^\circ$ \\
$^{16}$O & 600       &  7.6$^\circ$ & 1.4$^\circ$ \\
$^{16}$O & 1000      &  7.6$^\circ$ & 1.4$^\circ$ \\
$^{20}$Ne & 290      &  4.8$^\circ$ & 1.6$^\circ$ \\
$^{20}$Ne & 400      &  7.0$^\circ$ & 1.8$^\circ$ \\
$^{20}$Ne & 600      &  7.0$^\circ$ & 2.5$^\circ$, 1.7$^\circ$ \\
$^{24}$Mg & 400      &  9.5$^\circ$ & 2.0$^\circ$
\label{tab1}
\end{tabular}
\end{ruledtabular}
\end{table*}

The variation in large acceptance angles is not significant. In all cases, acceptance of fragments
with $Z$ $\geq$ $Z_{\text{beam}}$/2 is calculated to be at least 99{\%}. The variations in small acceptance
angles are more significant, as the measured spectra are found to be quite sensitive to both angle
and beam energy.

\subsection{Correction Factors}

The raw counts of events by species must be corrected for various effects before cross sections can be computed. 
The correction factors have been described in detail previously \cite{Zeitlin2001,Zeitlin2008}. The corrections 
are summarized in Table II. All have been applied to the cross sections presented below.

\begin{table*}
\caption{\label{table:tab2}List of correction factors applied to charge spectra.}
\begin{ruledtabular}
\begin{tabular}
%{|p{55pt}|p{58pt}|p{64pt}|p{64pt}|p{70pt}|p{70pt}|p{70pt}|}
{l l l}
Physical Effect & Effect on Spectra & Correction Estimation Method\\
\hline
Charge-changing interactions in & Loss of primary ions and       &  Target-out data \\
air gaps, dead layers, etc.     & increased numbers of fragments & \\
Multiple interactions in target & Shifts of fragment distributions & Monte Carlo simulation \cite{Zeitlin1996} \\ 
Charge-changing interactions in & $Z$-dependent detection efficiency & Geometric cross section model \\
the detector stack              &                                  & 
\label{tab2}
\end{tabular}
\end{ruledtabular}
\end{table*}

The magnitude of a given correction depends on the depth of the target (due to secondary and higher-order
interactions in the target) and the configuration of the silicon stack in a particular run. Corrections are 
smallest for the case of large-acceptance detectors close to a thin target. The correction factors have associated 
uncertainties that are taken into account when estimating systematic errors. Of the three effects listed in
Table II, the corrections for multiple interactions in the target are typically the largest, particularly
for the lower-mass targets (CH$_2$ and C) where even modest depths on the order of 3 g cm$^{-2}$ cause $\approx$
20\% of the beam ions to fragment, leading to $\approx$ 10\% corrections for the heaviest fragments.

Projectile fragments generally receive fairly small transverse momenta in the collisions, so that their angular
distributions are strongly forward-peaked. Multiple scattering angles are small at the energies considered here,
and can be ignored in determining large-acceptance cross sections. Because, as mentioned above, the large-acceptance 
data also require the smallest corrections for losses in the detectors and intervening materials, we use them
to obtain the charge-changing cross sections ($\sigma_{cc}$) and fragment cross sections for all resolvable species.
As can be seen in Fig. 2 above, where only peaks for B and C fragments are clear, this can be as little as two 
fragment species.

\section{Systematic Errors}

The statistical errors are generally small in these experiments, but the systematic uncertainties contain several 
contributions and typically dominate the total. When sufficient beam time is available, we take at least two runs 
with the same beam ion/energy/target combinations and vary the depth of the target. Cross sections obtained at 
different depths of the same material must, after corrections, be equal. The variations in the cross sections 
obtained this way are a good measure of the overall systematic uncertainties. When we do not have multiple data sets 
to combine, we associate a conservatively-large error with our data selection cuts and propagate the uncertainties
into all cross sections, as described in the next section.

As a practical matter, one can only obtain reasonable fragment statistics by using targets whose depths represent at least 
a few percent of an interaction length. With high-$Z$ targets like Sn and Pb, ionization energy losses per unit interaction 
length are large compared to low-$Z$ materials. In order to keep the beam energy approximately constant throughout the depth, 
high-$Z$ targets must therefore be kept thin, yielding poorer fragment statistics and larger relative systematic errors on 
the cross sections compared to lower-mass targets such as C and Al, for reasons that will become apparent in the following.

\subsection{Uncertainties in the Charge-Changing Cross Sections}

The definition of the cut contour that defines the surviving primaries and heaviest fragments is the single largest source 
of systematic error in these measurements. The most difficult contours to draw are those for runs with either no target 
or a very thin target, because in these cases the tail of the primary distribution on the low side can be difficult 
or impossible to distinguish from $\Delta Z$ = 1 events. Even with thicker targets, there is always at least a small
number of ambiguous events that fall between the clusters of primary ions and the highest-$Z$ fragments. To account for 
the cut contour uncertainty, a systematic error $\delta f$ is assigned to the fraction $f$ of surviving primaries. The magnitude 
of $\delta f$ is determined by repeatedly drawing the contours and examining the results. Typical variations in $f$ 
are 0.005 or smaller. In order to get a better sense of the magnitude of this uncertainty as it propagates into the 
charge-changing cross section, $\sigma_{cc}$, consider that the cross section scales with $ln(f)$. For thin targets, 
it is approximately true that $\sigma_{cc} \propto (1 - f)$, so that 
$\delta \sigma_{cc} / \sigma_{cc}  \propto \delta f / (1-f)$. 
Since $\delta f$ is found to be more or less constant for a given experiment, $\delta \sigma_{cc}$ is 
largest when $f$ approaches 1, corresponding to the thin-target case.

The preceding argument would appear to favor the use of thicker targets. However, that is only true to the extent
that cross sections can be approximated as energy independent. As target depth increases, so too does the range of
energies at which the primaries can interact. This can obscure subtleties in the energy dependences of either $\sigma_{cc}$ 
or the fragment production cross sections.

The model used to correct for losses due to nuclear interactions in the detectors is estimated to contribute 
on the order of $\pm$ 1\% relative error to the systematic uncertainty in $\sigma_{cc}$, and the accuracy of the 
target areal density measurements is also estimated at $\pm$ 1{\%}. For thin targets, the uncertainty associated with 
the cut contour dominates the quadrature sum, but these smaller contributions can be important for thicker targets. 
Determination of the systematic error on a given $\sigma_{cc}$ is made by combining the results from multiple targets. 
In combining data sets, the weighted average and a $\chi^2$ are computed. The error on each measurement is initially 
set equal to that arising from the definition of the primary selection cut contour. If this initial $\chi^2$ is found 
to be greater than 1.0 per degree of freedom (the number of data sets minus one), then an additional systematic error 
is added in quadrature to the starting errors of each individual measurement, and incremented upward in steps of 0.1\% 
relative error until we achieve the desired $\chi^2$ result ($<$ 1 per degree of freedom). In practice, these extra contributions 
are often not required since the estimates of the cut contour uncertainties are conservatively large and the initial 
calculation of $\chi^2$ typically yields a value $<$ 1 per degree of freedom.

\subsection{Uncertainties on Fragment Cross Sections at Large Acceptance}

For each data set, the uncertainty on the charge-changing cross section is propagated into the fragment cross sections, 
and added in quadrature to the statistical errors. Statistical errors are much more significant for fragments than 
for primaries. When data sets are combined, we again allow for the addition in quadrature of additional systematic 
uncertainties sufficient to bring the total $\chi^2$ for combining data (summed over all fragment species) to less than 
1.0 per degree of freedom. We find it is common that non-zero addition uncertainties are required at this step, in contrast
to what is seen when combining data sets to obtain charge-changing cross sections. We believe the major contributions that
these \textit{ad hoc} additions are covering are the uncertainties associated with ambiguities in (1) the cut contour that 
defines the sample of the heaviest fragments, and (2) the counting of events in the ``valleys'' between fragment peaks. 

\subsection{Additional Uncertainties on Fragment Cross Sections at Small Acceptance}

The uncertainties defined above contribute to the light-fragment cross sections. In addition, the acceptance corrections 
that are made to account for the fragment angular spreads contribute to the overall uncertainty. Previously, the relative 
errors associated with these corrections were found to be about $\pm$ 5-6{\%} \cite{Zeitlin2008}. This accounts for reasonable
variations in the size of the beam and its divergence, the model parameter $\sigma_0$, which controls the widths of the 
momentum distributions, and the exact active areas of the detectors. Here, a $\pm$ 6\% uncertainty is added in quadrature 
with the other uncertainties.

\begin{figure*}
\includegraphics[width=5.0in]{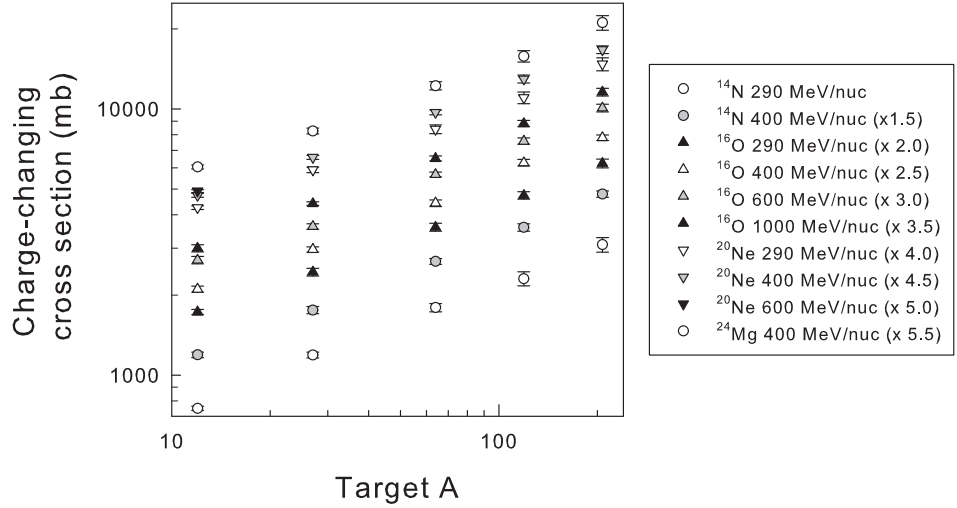}
\caption{\label{fig:fig3}Scaled charge-changing cross sections for all beams reported here, for targets from
carbon to lead.}
\end{figure*}

\section{Charge-Changing Cross Sections}

Table III shows the charge-changing cross section results. The same results are shown in Figure 3, but without
the hydrogen-target cross sections in order to keep the y-axis readable, and with scaling factors applied as
indicated in the figure legend, again for reasons of readability. NUCFRG2 and PHITS predictions are also shown 
in Table III. The model comparisons are discussed in the following section.

In general, the charge-changing cross sections shown here for carbon and heavier targets follow the same trends 
reported in our previous work with other beam ions in the same energy range: there is little or no energy dependence, 
and a simple geometrical model of overlapping spheres with a transparency term reproduces the data well. The only
apparent oddity in the results in Table III is the fact that the two hydrogen-target cross sections for $^{14}$N beams
at 290 and 400 MeV/nucleon are larger than those obtained with $^{16}$O beams at the same energies. However, that
refers only to the central values; taking into account the uncertainties, the data are almost equally consistent
with the hypothesis that the $^{16}$O cross sections are larger, as one would expect from simple geometry.
The only other unexpected trend in these data is the slight decrease of the $^{20}$Ne cross sections for aluminum 
and heavier targets at 600 MeV/nucleon compared to the 400 MeV/nucleon data. The effect is only slightly larger than 
the combined one-sigma uncertainties and we do not believe it has any physical significance.

\subsection{Comparison to Previous Measurements}
The most relevant published data to which we compare are drawn from a large number of charge-changing cross sections 
\cite{Webber1990a} and fragment production cross sections \cite{Webber1990b} published by Webber et al. We have made 
several previous comparisons to these data; in many cases, the differences are larger than the uncertainties. For
instance, in Ref. \cite{Zeitlin2007a}, we found discrepancies on the order of 5\% between our data and Webber's
when both carbon-target and hydrogen-target data were considered. Larger discrepancies were seen in several
instances where fragment production cross sections were compared. The situation is similar with the present
data sets. Table IV shows results for seven of the ten data sets analyzed here; the seven were
selected for similarities in beam energies. Agreement between the hydrogen-target cross sections is generally good, 
but it is not good for several of the carbon-target results. Specifically, the cross sections for oxygen beams on
carbon targets do not agree well, and the results for $^{20}$Ne on carbon at 400 MeV/nucleon differ by about 8{\%}. 
The discrepancies are all well beyond the stated uncertainties from either group. An additional independent measurement 
might be useful in these cases.

\begin{table*}
\caption{\label{table:tab3}Measured charge-changing cross sections and uncertainties, along with
NUCFRG2 and PHITS predictions. Listed energies are for the extracted beams.}
\begin{ruledtabular}
\begin{tabular}
%{|p{55pt}|p{58pt}|p{64pt}|p{64pt}|p{70pt}|p{70pt}|p{70pt}|}
{c c c c c c c}
Beam Ion, Energy& H target& C target& Al target& Cu target& Sn target& Pb target\\
\hline
$^{14}$N, 290& 229 $\pm $ 11 & 750 $\pm $ 14 & 1190 $\pm $ 31 & 1794 $\pm $ 64 & 2303 $\pm $ 140 & 3092 $\pm $ 197 \\
NUCFRG2      & 223           & 853           & 1198           & 1741           & 2353            & 3046 \\
%NUCFRG2      & 234           & 813           & 1154           & 1710           & 2306            & 3049 \\
PHITS        & 231           & 782           & 1149           & 1767           & 2314            & 3016 \\
\hline
$^{14}$N, 400& 236 $\pm $ 11 & 795 $\pm $ 19 & 1170 $\pm $ 36 & 1784 $\pm $ 45 & 2394 $\pm $ 78  & 3193 $\pm $ 88 \\
NUCFRG2      & 232           & 866           & 1215           & 1764           & 2384            & 3090 \\
%NUCFRG2      & 246           & 823           & 1166           & 1723           & 2321            & 3067 \\
PHITS        & 246           & 776           & 1136           & 1718           & 2227            & 2985 \\
\hline
$^{16}$O, 290& 219 $\pm $ 13 & 863 $\pm $ 20 & 1219 $\pm $ 41 & 1798 $\pm $ 60 & 2369 $\pm $ 74  & 3125 $\pm $ 118 \\
NUCFRG2      & 226           & 923           & 1287           & 1857           & 2495            & 3217 \\
%NUCFRG2      & 238           & 860           & 1210           & 1779           & 2387            & 3142 \\
PHITS        & 223           & 832           & 1220           & 1859           & 2436            & 3231 \\
\hline
$^{16}$O, 400& 220 $\pm $ 17 & 842 $\pm $ 22 & 1186 $\pm $ 27 & 1772 $\pm $ 51 & 2512 $\pm $ 72  & 3121 $\pm $  74 \\
NUCFRG2      & 237           & 937           & 1304           & 1881           & 2528            & 3264 \\
%NUCFRG2      & 251           & 871           & 1222           & 1793           & 2403            & 3162 \\
PHITS        & 245           & 827           & 1208           & 1840           & 2434            & 3217 \\
\hline
$^{16}$O, 600& 264 $\pm $ 17 & 902 $\pm $ 28 & 1206 $\pm $ 29 & 1892 $\pm $ 40 & 2524 $\pm $ 69  & 3366 $\pm $ 112 \\
NUCFRG2      & 259           & 976           & 1351           & 1940           & 2603            & 3359 \\
%NUCFRG2      & 266           & 910           & 1266           & 1847           & 2467            & 3239 \\
PHITS        & 269           & 856           & 1211           & 1818           & 2404            & 3224 \\
\hline
$^{16}$O, 1000& 276 $\pm $ 16 & 856 $\pm $ 26 & 1260 $\pm $ 19 & 1864 $\pm $ 40 & 2518 $\pm $  69 & 3307 $\pm $ 112 \\
NUCFRG2       & 285           & 1024          & 1408           & 2011           & 2692            & 3479  \\
%NUCFRG2       & 275           & 959           & 1324           & 1916           & 2551            & 3345  \\
PHITS         & 280           & 937           & 1211           & 1771           & 2312            & 3116 \\
\hline
$^{20}$Ne, 290& 272 $\pm $ 16 & 1050 $\pm $ 21 & 1445 $\pm $ 45 & 2043 $\pm $ 99  & 2807 $\pm $ 142 & 3556 $\pm $ 260 \\
NUCFRG2       & 287           & 1062           & 1462           & 2083            & 2772            & 3548 \\
%NUCFRG2       & 299           & 949            & 1315           & 1907            & 2538            & 3318 \\
PHITS         & 281           & 998            & 1420           & 2107            & 2776            & 3543 \\
\hline
$^{20}$Ne, 400& 311 $\pm $ 15 & 1034 $\pm $ 21 & 1438 $\pm $ 26 & 2140 $\pm $ 47 & 2764 $\pm $ 99 & 3555 $\pm $  129 \\
NUCFRG2       & 298           & 1078           & 1481           & 2109           & 2810           & 3603 \\
%NUCFRG2       & 314           & 960            & 1327           & 1922           & 2556           & 3342 \\
PHITS         & 305           & 983            & 1416           & 2092           & 2723           & 3538 \\
\hline
$^{20}$Ne, 600& 319 $\pm $ 13 & 986 $\pm $ 14 & 1349 $\pm $ 30 & 1993 $\pm $ 35 & 2572 $\pm $ 75 & 3407$\pm $  73 \\
NUCFRG2       & 321           & 1119          & 1532           & 2173           & 2892           & 3711 \\
%NUCFRG2       & 331           & 1000          & 1374           & 1979           & 2625           & 3429 \\
PHITS         & 326           & 1018          & 1422           & 2098           & 2758           & 3580 \\
\hline
$^{24}$Mg, 400& 328 $\pm $ 18 & 1028 $\pm $ 18 & 1480 $\pm $ 38 & 2244 $\pm $ 79 & 2794 $\pm $ 142 & 3727 $\pm $ 239 \\
NUCFRG2       & 315           & 1124           & 1547           & 2194           & 2935            & 3761 \\
%NUCFRG2       & 349           & 1044           & 1427           & 2043           & 2698            & 3509 \\
PHITS         & 313           &  998           & 1428           & 2117           & 2808            & 3670
\label{tab3}
\end{tabular}
\end{ruledtabular}
\end{table*}

Since both groups used polyethylene and carbon targets to obtain the hydrogen-target results, and those are in
better agreement, it must also be the case that the polyethylene-target results have a level of disagreement similar 
to that seen for the carbon targets. The agreement for hydrogen is to some extent a product of the cancellation of errors 
in the subtraction.

In the results shown here, the level of agreement between our experimental results and those of Webber et al. is
not especially good. If errors were correctly estimated in all experiments, we would expect to see very few 
values of $\chi^2$ as large as those seen in Table IV (i.e., $>$ 5 for one degree of freedom). This comment
pertains to several previously-published data points as well. In some instances, there are additional data from 
other groups that can be used for additional comparisons. Our charge-changing cross sections are generally 
in good agreement with those measured using plastic nuclear track detectors \cite{Iancu2005, Flesch2001}.

\subsection{Charge-changing Cross Sections Compared to Model Predictions}
A discussion of the physics content of the various models used here is given below in Section IX.A.

As in earlier work, we compare the charge-changing cross section data to several models, including a simple formula
given by $\sigma_{cc}=10\pi r_0^2(A_t^{1/3}+A_p^{1/3}-b)^2$, 
allowing $b$ (dimensionless) and $r_0$ (in units of fm) to vary. Where $A_t$ and $A_p$ are the atomic mass numbers of the 
target and projectile, respectively.  For each charge-changing cross section and each model, we
calculated the ratio of the predicted cross section to the measured. We then averaged the results over all beam ion/energy
combinations for a particular target species. We then examined the aggregate results, that is, the results for all targets
combined. Tuning the parameters to values of $b$ = 0.938 and $r_0$ = 1.375 fm yields results that, using this averaging 
method, agree with the data to better than 0.3{\%}. (For the 600 MeV/nucleon $^{20}$Ne data reported earlier, we found 
good agreement with the data for $b =$ 0.86 and $r_0$ = 1.34 fm.) The near-perfect average agreement is not particularly 
significant -- the standard deviation relative to the mean is a better measure of typical accuracy over multiple beam ions, 
energies, and targets. 

The averaged results are shown in Table V, for this simple form and for NUCFRG2 and PHITS as well. Values below 1.0 
indicate that the predicted cross sections are smaller than the measured values. The simple, energy-independent model and PHITS
yield the same value (0.041) for the figure of merit, whereas NUCFRG2 gives a somewhat higher value (0.050), indicating
more variance between the data and NUCFRG2 than for the other models. If we exclude the cross sections obtained with 290
MeV/nucleon beams, the simple model yields an even better figure of merit, 0.035. No similar improvement is seen if the 
290 MeV/nucleon data are excluded from the PHITS and NUCFRG2 comparisons. Thus, the simple energy independent model can, 
at least for this very limited range of beam ions and energies, be tuned to a high degree of accuracy. It seems likely 
that other parameter values can be found that would give better agreement with other data sets (higher-energy and/or 
higher-mass beams); thus it may be reasonable to treat one or both of these parameters as energy dependent to obtain an 
empirical fit across a greater range of data.

Some trends are apparent in the averages by target. For NUCFRG2, the ratios of predicted to measured cross
sections are significantly greater than 1 for carbon and aluminum targets, with values closer to 1.0 found for Cu, Sn,
and Pb targets. NUCFRG2 also does well, on average, for hydrogen targets. The lack of agreement with the data for C and
Al targets is potentially important, since the beam ions studied here are relatively abundant among GCR heavy ions, and 
both C and Al are important shielding materials in space. For PHITS, the averages are highly consistent for C, Al, and Cu, 
then decrease slightly for Sn and Cu. 

The comparatively small spread in the ratios found for the simple model does not hold when this formula with these 
same parameter values is applied to other beam ion/target combinations. When we average over our published data for 
beams from 290 MeV/nucleon $^{12}$C to 1000 MeV/nucleon $^{56}$Fe, the average remains close to 1.0 (0.994), but the 
standard deviation increases to about 6.4{\%}, and some discrepancies appear that approach a 10{\%} level of disagreement. 
Many of these data points were obtained at higher energies than the data presented here.

In the following we will compare fragment production cross sections not only with PHITS and NUCFRG2, as above, but
also with the predictions of the Los Alamos version of the Quark Gluon String Model (LAQGSM) \cite{Mashnik2008a,Mashnik2008b}. 
In the output generated by the LAQGSM code, charge-changing cross sections are not provided. Rather, cross sections are 
computed as elastic or inelastic. This does not allow for comparisons to the data above, since the inelastic and
charge-changing cross sections do not correspond to one another. The inelastic cross section contains the charge-changing 
cross sections plus contributions from neutron-stripping reactions, and is therefore larger than the charge-changing cross 
section alone. The elastic cross section cannot be measured in our experiments. Therefore, meaningful comparisons of the
charge-changing cross section data to LAQGSM predictions are not possible given the current state of the output from the code. 
But as we will show below, valuable comparisons can be made for fragment production cross sections.

\begin{table*}
\caption{\label{table:tab4}Charge-changing cross sections for hydrogen and carbon targets as reported here
and as measured by Webber et al. \cite{Webber1990a}. Listed beam energies are at the target centers. The
$\chi^2$ values are all for one degree of freedom.}
\begin{ruledtabular}
\begin{tabular}
%{|p{55pt}|p{58pt}|p{64pt}|p{64pt}|p{70pt}|p{70pt}|p{70pt}|}
{c c c c c c c c}
Beam Ion & Energy        & H target  & H target              & $\chi^2$ for & C target  & C target & $\chi^2$ for \\
         & (MeV/nucleon) & This work & Ref. \cite{Webber1990a} & agreement  & This work & Ref. \cite{Webber1990a}& agreement\\
\hline
N        & 375, 516 & 236 $\pm$ 12 & 227 $\pm$ 6 & 0.5 & 795 $\pm$ 19 & 796 $\pm$ 8 & 0.00 \\
O        & 375, 441 & 224 $\pm$ 17 & 232 $\pm$ 6 & 0.2 & 872 $\pm$ 22 & 794 $\pm$ 8 & 11.1 \\
O        & 578, 491 & 264 $\pm$ 17 & 247 $\pm$ 6 & 0.9 & 902 $\pm$ 28 & 823 $\pm$ 8 & 7.4 \\
O        & 980, 903 & 276 $\pm$ 16 & 248 $\pm$ 6 & 2.7 & 856 $\pm$ 26 & 813 $\pm$ 8 & 2.5 \\
Ne       & 375, 468 & 311 $\pm$ 15 & 298 $\pm$ 7 & 0.6 & 1034 $\pm$ 21 & 951 $\pm$ 10 & 12.7 \\
Ne       & 572, 599 & 311 $\pm$ 14 & 319 $\pm$ 8 & 0.2 & 984 $\pm$ 13 & 980 $\pm$ 10 & 0.1 \\
Mg       & 358, 309 & 328 $\pm$ 18 & 310 $\pm$ 8 & 0.8  & 1028 $\pm$ 19 & 1097 $\pm$ 11 & 9.9
\label{tab4}
\end{tabular}
\end{ruledtabular}
\end{table*}

\begin{table*}
\caption{\label{table:tab5}Averaged ratios of predicted charge-changing cross sections to measured cross sections.
There are two entries in each cell, the first being the averaged ratio and the second being the standard deviation.
The simple model is energy-independent, while NUCFRG2 and PHITS contain slight energy dependence, which is weak
except for hydrogen-target cross sections. In the bottom row, ratios have been re-computed for the simple model
excluding data from the 290 MeV/nucleon beams.}
\begin{ruledtabular}
\begin{tabular}
%{|p{55pt}|p{58pt}|p{64pt}|p{64pt}|p{70pt}|p{70pt}|p{70pt}|}
{c c c c c c c c}
Model & H target& C target& Al target& Cu target& Sn target& Pb target& Combined\\
\hline
NUCFRG2    & 1.006, 0.041 & 1.097, 0.052 & 1.066, 0.048 & 1.023, 0.043 & 1.036, 0.041 & 1.019, 0.036 & 1.041, 0.050 \\
PHITS      & 1.021, 0.041 & 0.991, 0.049 & 0.991, 0.029 & 0.994, 0.041 & 0.985, 0.046 & 0.982, 0.034 & 0.994, 0.041 \\
Simple     & n/a          & 1.019, 0.055 & 0.995, 0.037 & 0.980, 0.036 & 1.003, 0.036 & 1.010, 0.033 & 1.000, 0.041 \\
Simple$^*$ & n/a          & 1.015, 0.041 & 1.000, 0.036 & 0.975, 0.040 & 0.997, 0.022 & 1.010, 0.030 & 1.000, 0.035
\label{tab5}
\end{tabular}
\end{ruledtabular}
\end{table*}

\section{Fragment Production Cross Sections at Large Acceptance}
Fragment production cross sections are presented in two parts. We first discuss those obtained at large acceptance,
where it is typically possible only to measure fragment species with (approximately) $Z_{\text{frag}} \geq Z_{\text{beam}}/2$. 
These data are readily compared to model predictions, and in some instances, we are also able to compare to 
previously published data from Webber et al. \cite{Webber1990b} for hydrogen and carbon targets. In the second part 
of the discussion, we present cross sections for lighter fragments measured with the small-acceptance detectors. 
These cross sections are (with a few exceptions) corrected for acceptance losses using a previously-described
method \cite{Zeitlin2001, Zeitlin2007b} based on Goldhaber's model of nucleon momentum distribution within
the nucleus \cite{goldhaber}, subsequently modified by Tripathi and Townsend \cite{triptowns}. Complications
in interpreting these light-fragment cross sections are discussed below.

Direct comparisons of the results obtained at large acceptance can be made to a few previous measurements, particularly 
those of Webber et al. \cite{Webber1990b}. We return to this point below in Section IX.E. 
No measurements comparable to our small-acceptance results are in the literature to our knowledge.

\subsection{Nitrogen Beams}
For the 290 MeV/nucleon $^{14}$N data, we are able to resolve charge 4 peaks, whereas in the 400 MeV/nucleon $^{14}$N 
spectra we can only see fragment peaks for charges 5 and 6. This is probably due to the greater forward-focusing of
non-leading fragments at the higher energy, resulting in a higher average detected multiplicity. Table VI shows the
results for nitrogen beams at two energies. The charge 4 cross sections for the 290 MeV/nucleon beam are presented 
below, with the small-acceptance results. The results are quite similar at these two energies. The $\Delta Z$ = 1
cross sections appear systematically smaller by 8-10{\%} at 400 MeV/nucleon for all but the H and Pb targets. (For
the Pb target, the large error on the 290 MeV/nucleon data point precludes drawing any such conclusion.) This is
qualitatively consistent with trends we have observed in other data sets, in which the cross sections for the smallest 
charge changes decrease slightly with energy while those for larger charge changes are more constant or even increase.
For boron fragments (charge 5) no significant differences are seen between the two energies. The boron production
cross sections are roughly a factor of two lower than those for carbon production. For charge 4 fragments, the
pattern seen for charge 6 fragments repeats; that is, for C, Al, Cu, and Sn targets, the 400 MeV/nucleon cross
sections are systematically smaller by 10-25\% than those measured with the 290 MeV/nucleon beam. For both
charge 4 and 6 fragments, the Pb-target uncertainties are too large to say whether or not they are consistent
with the trend observed for C through Sn targets.

In the last two rows of Table VI, we show the cross section for events in which the charge of the leading 
particle is
less than 4. Using the large acceptance detectors, there is no more information to be gleaned from these
events. However, it is clear that these cross sections increase monotonically with target mass as a share of 
the charge changing cross section. These cross sections generally increase with increasing beam energy. The
trend is not statistically significant in the $^{14}$N data owing to the small difference in the two beam
energies in these data, but it is seen clearly in other data sets.

\begin{table*}
\caption{\label{table:tab6}Fragment production cross section for $^{14}$N beams on elemental targets. Beam energies
at extraction are shown, in units of MeV/nucleon. All cross sections are in millibarns.}
\begin{ruledtabular}
\begin{tabular}
{c c c c c c c c}
$Z_{\text{frag}}$ & $E_{\text{beam}}$ & H          & C           & Al          & Cu           & Sn           & Pb \\
\hline
6          & 290        & 82 $\pm$ 8 & 162 $\pm$ 3 & 221 $\pm$ 7 & 283 $\pm$ 11 & 319 $\pm$ 22 & 389 $\pm$ 28\\
6          & 400        & 83 $\pm$ 4 & 150 $\pm$ 4 & 195 $\pm$ 8 & 260 $\pm$ ~8 & 312 $\pm$ 13 & 415 $\pm$ 15 \\
\hline
5          & 290        & 29 $\pm$ 3 &  72 $\pm$ 2 &  98 $\pm$ 3 & 132 $\pm$ 6  & 152 $\pm$ 11 & 176 $\pm$ 13 \\
5          & 400        & 27 $\pm$ 2 &  75 $\pm$ 2 &  96 $\pm$ 4 & 127 $\pm$ 4  & 154 $\pm$ ~7 & 176 $\pm$ ~7 \\
\hline
4          & 290        & 14 $\pm$ 2 & 48 $\pm$ 2 &  67 $\pm$ 3  &  97 $\pm$ 5  & 115 $\pm$ 9  & 133 $\pm$ 11 \\
4          & 400        & 15 $\pm$ 2 & 43 $\pm$ 2 &  59 $\pm$ 4  &  81 $\pm$ 5  & ~92 $\pm$ 7  & 130 $\pm$ ~9  \\
\hline
$<$4       & 290      &104 $\pm$ 11 & 468 $\pm$ 14 & 804 $\pm$ 31 & 1282 $\pm$ 64 & 1717 $\pm$ 140 & 2394 $\pm$ 197 \\
$<$4       & 400      &111 $\pm$ 12 & 526 $\pm$ 19 & 820 $\pm$ 36 & 1316 $\pm$ 45 & 1836 $\pm$ ~78 & 2472 $\pm$ ~88 
\label{tab6}
\end{tabular}
\end{ruledtabular}
\end{table*}

\subsection{Oxygen Beams}
For the four $^{16}$O beams studied, fragment charges from 5 to 7 can be resolved at large acceptance. 
The results are shown in Table VII. As in the $^{14}$N data above, the $\Delta Z$ = 1 cross sections
(charge 7) tend to decrease with increasing beam energy for all targets except H and Pb. This is also predominantly
the case for $\Delta Z$ = 2 and $\Delta Z$ = 3, though there are a few exceptions to the general trend. Since the
charge-changing cross sections are, except for hydrogen targets, roughly constant, the cross sections for the 
category $Z$ $<$ 5 increase with increasing beam energy. This is a modest effect, on the order of 10 to 15\% for
carbon and heavier targets. The hydrogen target data show somewhat different behavior: the charge-changing 
cross section increases with energy, and the cross section in the $Z$ $<$ 5 category also increases as energy 
increases, while the cross sections for $\Delta Z$ = 1, 2, and 3 are approximately independent of energy in this range.

The odd-$Z$ even-$Z$ effect, discussed in detail elsewhere \cite{Zeitlin2008}, can be seen here: the production cross sections 
for $\Delta Z$ = 2 are higher than the corresponding cross sections for $\Delta Z$ = 1 in all cases for H, C, Al, and 
Cu targets, but not for Sn and Pb targets. A similar trend was seen in silicon-beam data \cite{Zeitlin2007a}. However, 
with heavier beams ($^{40}$Ar, $^{48}$Ti, and $^{56}$Fe), the opposite is true. As was the case for the $^{14}$N beams, 
the boron production cross sections are 50 to 60{\%} as large as those for carbon production.

\begin{table*}
\caption{\label{table:tab7}Fragment production cross section for $^{16}$O beams on elemental targets. Beam energies
at extraction are shown, in units of MeV/nucleon. All cross sections are in millibarns.}
\begin{ruledtabular}
\begin{tabular}
{c c c c c c c c}
$Z_{\text{frag}}$ & $E_{\text{beam}}$ & H          & C           & Al          & Cu           & Sn           & Pb \\
\hline
7          & 290        & 59 $\pm$ 3 & 137 $\pm$ 4 & 169 $\pm$ 7 & 204 $\pm$ 8  & 247 $\pm$ 10 & 297 $\pm$ 15 \\
7          & 400        & 58 $\pm$ 4 & 124 $\pm$ 4 & 145 $\pm$ 5 & 187 $\pm$ 8  & 244 $\pm$ 11 & 269 $\pm$ 10 \\
7          & 600        & 65 $\pm$ 4 & 120 $\pm$ 5 & 138 $\pm$ 4 & 188 $\pm$ 6  & 240 $\pm$ 10 & 285 $\pm$ 14 \\
7          & 1000       & 59 $\pm$ 3 & 105 $\pm$ 4 & 131 $\pm$ 3 & 167 $\pm$ 5  & 210 $\pm$ ~8 & 270 $\pm$ 13 \\
\hline
6          & 290        & 61 $\pm$ 4 & 148 $\pm$ 4 & 181 $\pm$ 7 & 225 $\pm$ 8  & 258 $\pm$ 11 & 305 $\pm$ 15 \\
6          & 400        & 60 $\pm$ 4 & 131 $\pm$ 5 & 159 $\pm$ 5 & 209 $\pm$ 8  & 241 $\pm$ 12 & 257 $\pm$ 10 \\
6          & 600        & 71 $\pm$ 4 & 131 $\pm$ 5 & 154 $\pm$ 5 & 206 $\pm$ 7  & 248 $\pm$ 10 & 279 $\pm$ 14 \\
6          & 1000       & 63 $\pm$ 3 & 121 $\pm$ 4 & 153 $\pm$ 4 & 199 $\pm$ 6  & 225 $\pm$ ~8 & 281 $\pm$ 13 \\
\hline
5          & 290        & 23 $\pm$ 2 & 83 $\pm$ 2 & 101 $\pm$ 4 & 143 $\pm$ 5  & 157 $\pm$ 13 & 201 $\pm$ 11 \\
5          & 400        & 25 $\pm$ 2 & 70 $\pm$ 3 & ~86 $\pm$ 3 & ~99 $\pm$ 5  & 128 $\pm$ ~7 & 149 $\pm$ ~6 \\
5          & 600        & 25 $\pm$ 2 & 72 $\pm$ 3 & ~76 $\pm$ 3 & 112 $\pm$ 4  & 123 $\pm$ ~6 & 168 $\pm$ ~9 \\
5          & 1000       & 29 $\pm$ 2 & 56 $\pm$ 2 & ~72 $\pm$ 2 & ~91 $\pm$ 3  & 114 $\pm$ ~5 & 130 $\pm$ ~7 \\
\hline
$<$5       & 290     & 76 $\pm$ 13 & 495 $\pm$ 20 & 768 $\pm$ 41 & 1226 $\pm$ 60 & 1707 $\pm$ 74 & 2322 $\pm$ 118 \\
$<$5       & 400     & 77 $\pm$ 17 & 517 $\pm$ 22 & 796 $\pm$ 27 & 1277 $\pm$ 51 & 1899 $\pm$ 72 & 2446 $\pm$ ~74 \\
$<$5       & 600     &103 $\pm$ 13 & 579 $\pm$ 28 & 838 $\pm$ 29 & 1386 $\pm$ 40 & 1913 $\pm$ 69 & 2634 $\pm$ 112 \\
$<$5       & 1000    &125 $\pm$ 16 & 574 $\pm$ 26 & 904 $\pm$ 19 & 1407 $\pm$ 40 & 1969 $\pm$ 69 & 2626 $\pm$ 112 
\label{tab7}
\end{tabular}
\end{ruledtabular}
\end{table*}

\subsection{Neon Beams}
Three $^{20}$Ne beam energies were studied. Fragment charges from 5 to 9 can be resolved at large acceptance.
The results are shown in Table VIII. The same four trends discussed above for N and O beams can be seen here:
even-Z fragment production cross sections are enhanced compared to those for odd $Z$'s; for a given fragment 
species and target, cross sections for the species that can be resolved decrease with increasing energy for 
carbon and heavier targets; boron production cross sections are again about a factor of 2 smaller 
than those for carbon production; and the cross section for the unresolved category increases with
increasing beam energy. 

We previously published results for 600 MeV/nucleon $^{20}$Ne on the same target materials \cite{Zeitlin2001}.
Those cross sections have been combined with an additional 600 MeV/nucleon data set obtained subsequently.
Our methods for combining data sets and estimating systematic errors have changed in the interim; the
results shown here fully incorporate the newer methodology.

\begin{table*}
\caption{\label{table:tab8}Fragment production cross section for $^{20}$Ne beams on elemental targets. Beam energies
at extraction are shown, in units of MeV/nucleon. All cross sections are in millibarns.}
\begin{ruledtabular}
\begin{tabular}
{c c c c c c c c}
$Z_{\text{frag}}$ & $E_{\text{beam}}$ & H          & C           & Al          & Cu           & Sn           & Pb \\
\hline
9          & 290        & 49 $\pm$ 3 & 109 $\pm$ 3 & 130 $\pm$ 5 & 175 $\pm$ 11 & 229 $\pm$ 14 & 308 $\pm$ 27 \\
9          & 400        & 47 $\pm$ 3 & 102 $\pm$ 3 & 120 $\pm$ 3 & 160 $\pm$ ~5 & 195 $\pm$ 12 & 236 $\pm$ 13 \\
9          & 600        & 53 $\pm$ 2 & ~84 $\pm$ 2 & 106 $\pm$ 3 & 144 $\pm$ ~4 & 177 $\pm$ ~7 & 235 $\pm$ ~8 \\
\hline
8          & 290        & 78 $\pm$ 4 & 163 $\pm$ 4 & 188 $\pm$ 7 & 230 $\pm$ 13 & 281 $\pm$ 17 & 304 $\pm$ 27 \\
8          & 400        & 81 $\pm$ 4 & 151 $\pm$ 4 & 178 $\pm$ 5 & 228 $\pm$ ~7 & 262 $\pm$ 15 & 318 $\pm$ 15 \\
8          & 600        & 75 $\pm$ 3 & 134 $\pm$ 3 & 158 $\pm$ 5 & 205 $\pm$ ~6 & 237 $\pm$ ~9 & 296 $\pm$ 10 \\
\hline
7          & 290        & 54 $\pm$ 3 & 128 $\pm$ 4 & 155 $\pm$ 6 & 191 $\pm$ 11 & 240 $\pm$ 15 & 275 $\pm$ 24 \\
7          & 400        & 55 $\pm$ 3 & 117 $\pm$ 3 & 144 $\pm$ 4 & 179 $\pm$ ~6 & 206 $\pm$ 12 & 241 $\pm$ 12 \\
7          & 600        & 55 $\pm$ 3 & 101 $\pm$ 2 & 123 $\pm$ 4 & 156 $\pm$ ~5 & 183 $\pm$ ~7 & 222 $\pm$ ~8 \\
\hline
6          & 290        & 48 $\pm$ 3 & 161 $\pm$ 4 & 190 $\pm$ 7 & 223 $\pm$ 13 & 272 $\pm$ 16 & 314 $\pm$ 27 \\
6          & 400        & 55 $\pm$ 3 & 144 $\pm$ 4 & 170 $\pm$ 5 & 222 $\pm$ ~7 & 258 $\pm$ 14 & 282 $\pm$ 13 \\
6          & 600        & 56 $\pm$ 3 & 124 $\pm$ 3 & 152 $\pm$ 5 & 200 $\pm$ ~6 & 244 $\pm$ ~9 & 271 $\pm$ ~9 \\
\hline
5          & 290        & 11 $\pm$ 4 & 75 $\pm$ 5 & 100 $\pm$ 4 & 137 $\pm$ ~8 & 151 $\pm$ 10 & 184 $\pm$ 18 \\
5          & 400        & 22 $\pm$ 2 & 79 $\pm$ 4 & ~98 $\pm$ 5 & 126 $\pm$ ~6 & 143 $\pm$ 15 & 161 $\pm$ ~8 \\
5          & 600        & 26 $\pm$ 2 & 65 $\pm$ 2 & ~81 $\pm$ 3 & 108 $\pm$ ~4 & 123 $\pm$ ~6 & 153 $\pm$ ~7 \\
\hline
$<$5       & 290        & 33 $\pm$ 16 & 414 $\pm$ 21 & 692 $\pm$ 45 & 1100 $\pm$ 99 & 1660 $\pm$ 142 & 2239 $\pm$ 260 \\
$<$5       & 400        & 51 $\pm$ 15 & 441 $\pm$ 21 & 728 $\pm$ 26 & 1225 $\pm$ 47 & 1700 $\pm$ ~99 & 2317 $\pm$ 129 \\
$<$5       & 600        & 52 $\pm$ 14 & 475 $\pm$ 13 & 708 $\pm$ 36 & 1205 $\pm$ 48 & 1727 $\pm$ ~66 & 2270 $\pm$ 62 
\label{tab8}
\end{tabular}
\end{ruledtabular}
\end{table*}

\subsection{$^{24}$Mg at 400 MeV/nucleon}
Data were obtained with a 400 MeV/nucleon $^{24}$Mg beam. Fragment charges from 6 to 11 can be resolved at large acceptance.
These results are shown in Table IX. Also shown in Table IX are the cross sections for boron fragments (charge 5), which were 
obtained at small acceptance; corrections have been applied.

\subsection{The Odd-Z Even-Z Effect}
For the $^{14}$N and $^{16}$O beams discussed above, cross sections for just three fragment species 
can be measured at large acceptance. This is not a sufficient number to allow us to make any statements
about the odd-$Z$/even-$Z$ effect. However, with Ne and Mg beams, we can explore
the magnitude of the effect in a quantitative way. Both beam ions have isospin $T_z = 0$. As in previous work 
\cite{Zeitlin2008}, we use the quantity $V(Z)$ defined by Iancu et al. \cite{Iancu2005}:
\[
V\left( {Z_f } \right)={2\sigma \left( {Z_f } \right)} \mathord{\left/ 
{\vphantom {{2\sigma \left( {Z_f } \right)} {\left[ {\sigma \left( {Z_f +1} 
\right)+\sigma \left( {Z_f -1} \right)} \right]}}} \right. 
\kern-\nulldelimiterspace} {\left[ {\sigma \left( {Z_f +1} \right)+\sigma 
\left( {Z_f -1} \right)} \right]}
\]
where $Z_f$ refers to fragments of charge $Z$. In Ref. \cite{Zeitlin2008}, we combined results for
C and Al targets. Here, examination of the $^{20}$Ne and $^{24}$Mg data shows no statistically
significant differences between C, Al, Cu, and Sn targets, so all are combined. Lead (Pb) targets
are excluded since results may be distorted by the contributions from electromagnetic dissociation,
which produces increases in the $\Delta Z$ = 1 (and perhaps $\Delta Z$ = 2) cross sections. We combine
the values of $V(Z_f)$ obtained for all odd-$Z$ fragments into a single weighted-average value, and 
similarly combine the results for all even-$Z$ fragments to get that weighted average, and take 
the ratio of the two to obtain a single value for a given beam ion and energy.

\begin{table*}
\caption{\label{table:tab9}Fragment production cross section for a $^{24}$Mg beam on elemental targets. The beam energy
was 400 MeV/nucleon at extraction. Cross sections are in millibarns.}
\begin{ruledtabular}
\begin{tabular}
{c c c c c c c c}
$Z_{\text{frag}}$ & H           & C            & Al           & Cu            & Sn            & Pb \\
\hline
11         & 72 $\pm$ 3  & 122 $\pm$ 3  & 154 $\pm$ 5  & 195 $\pm$ 8   & 232 $\pm$ 14  & 274 $\pm$ 22 \\
10         & 63 $\pm$ 3  & 112 $\pm$ 3  & 135 $\pm$ 4  & 176 $\pm$ 8   & 197 $\pm$ 12  & 250 $\pm$ 19 \\
9          & 27 $\pm$ 1  & ~56 $\pm$ 2  & ~64 $\pm$ 3  & ~86 $\pm$ 4   & ~96 $\pm$ ~7  & ~96 $\pm$ ~9 \\
8          & 56 $\pm$ 3  & 113 $\pm$ 3  & 136 $\pm$ 4  & 175 $\pm$ 7   & 181 $\pm$ 11  & 232 $\pm$ 18 \\
7          & 34 $\pm$ 2  & ~85 $\pm$ 2  & 114 $\pm$ 4  & 142 $\pm$ 6   & 180 $\pm$ 11  & 180 $\pm$ 14 \\
6          & 33 $\pm$ 2  & 113 $\pm$ 3  & 150 $\pm$ 5  & 200 $\pm$ 8   & 191 $\pm$ 12  & 280 $\pm$ 21 \\
5          & 12 $\pm$ 3  & ~53 $\pm$ 4  & ~71 $\pm$ 5  & ~97 $\pm$ 8   & 120 $\pm$ 12  & 132 $\pm$ 16 \\
$<$ 5      & 31 $\pm$ 18 & 374 $\pm$ 18 & 656 $\pm$ 38 & 1173 $\pm$ 79 & 1597 $\pm$ 79 & 2283 $\pm$ 239
\label{tab9}
\end{tabular}
\end{ruledtabular}
\end{table*}

For $^{20}$Ne beams, we can calculate $V(Z_f)$ only for charges 6, 7, and 8. We find overall ratios
of 1.71 $\pm$ 0.05, 1.73 $\pm$ 0.04, and 1.86 $\pm$ 0.04 for 290, 400, and 600 MeV/nucleon beam
energies, respectively. In Ref. \cite{Zeitlin2008}, we used this same method for $^{28}$Si beams
at extracted energies of 290, 400, 600, 800, and 1200 MeV/nucleon, and found that the even:odd 
$V(Z_f)$ ratio for C and Al targets increased slightly as beam energy increased in going from 290 to
400 and 600 MeV/nucleon, with ratios of 2.09 $\pm$ 0.09, 2.22 $\pm$ 0.08, and 2.32 $\pm$ 0.06 for the
three energies, respectively. The results for $^{20}$Ne show the same trend, although all ratios are 
smaller for the lighter beam. (Data points with $^{20}$Ne at higher energies would be of interest.)
For $^{24}$Mg at 400 MeV/nucleon, we find a ratio of 2.32 $\pm$ 0.05, compatible with that found 
for $^{28}$Si at the same beam energy.

\section{Small Acceptance Spectra: General Considerations}
In the following, we again present the results grouped by beam ion, similar to the way in which the 
large-acceptance results were presented above. However, here it will be seen that the differences
between corrected results at different beam energies are, in many instances, quite large, for reasons 
that are related to the variations in the acceptances (which depend on beam ion, energy, and experimental
configuration) and the associated correction factors. Before presenting the results, discussion of some 
broadly-applicable points is in order, as there are several caveats in the interpretation of these data.
The corrected light fragment cross sections presented here must be interpreted with considerable 
caution.

\subsection{Events with No Detected Fragments}
In some runs, the most populated region of the small-acceptance spectrum is that below the peak due to single
charge-1 particles. These are events in which either nothing at all was detected, or the detected particle(s)
deposited less than a minimum-ionizing charge 1 particle. These events may contain low-energy
electrons ($E$ $\approx$ 1 MeV), or Compton electrons produced in the detector from the traversal of an energetic
gamma-ray. These events are counted in the sense that they contribute to the charge-changing cross section,
as invariably some charge is recorded at large acceptance,
but no corrections are applied to this event category. We do not report these cross sections in the following.

\subsection{Non-leading Charge-1 Fragments}
Interpretation of the cross sections for the lightest species, H and He, must be approached with particular
caution. These fragments are copiously produced in the interactions studied here, but typically they are not 
likely to be the highest-$Z$ fragment detected in a given event, even at small acceptance. The experimental 
approach used here cannot fully account for all of these fragments and therefore the cross sections obtained
before acceptance corrections are applied represent the \textit{detected portion} of the overall production 
cross sections. Given the experimental setup, it is only feasible to measure the cross section for events 
in which either a H or He ion is the highest-charge fragment seen in the small-acceptance detector, and 
extrapolate these results using our angular acceptance model. Even with these corrections, we expect that 
large portions of the true production cross sections are missed. Consider the simplest case in which a single 
proton or deuteron is stripped from the projectile (i.e., $\Delta Z$ = 1), a hydrogen fragment must also be 
created; however, the probability that it is detected is small, for two reasons. First, the heavier fragment 
angular distribution is much more sharply forward-peaked than that of the H fragment. Conservation of transverse 
momentum dictates that even in those rare events where the heavier fragment is produced at a large enough angle 
to miss the small-acceptance detector, the H fragment or fragments (for $\Delta Z$ $>$ 1) must have balancing 
transverse momenta, so that they tend to be far outside the small acceptance. Secondly, the heavier fragment 
will virtually always be seen in the large-acceptance detector and has a very high probability to be seen in 
the small-acceptance detector, and with unsegmented detectors having limited resolution in $\Delta E$, a non-leading 
H fragment will typically not be detected even if it is within the acceptance because its contribution to the effective
detected charge is so small. For example, in oxygen beam data, we cannot distinguish between the detection of a 
nitrogen fragment alone, $Z_{\text{eff}}^2$ = 49, and coincidental detection of a nitrogen fragment and a proton, 
$Z_{\text{eff}}^2$ = 50.  (By $Z_{\text{eff}}$, we mean the ``effective" $Z$ seen in a particular detector pair; 
as a practical matter, this is proportional to the square root of $\Delta E$ and - for ions with equal velocities - also 
proportional to the square root of the sum of the charges squared. See Ref. \cite{Zeitlin2001} for additional explanation.)

\begin{figure}
\includegraphics[width=3.37in]{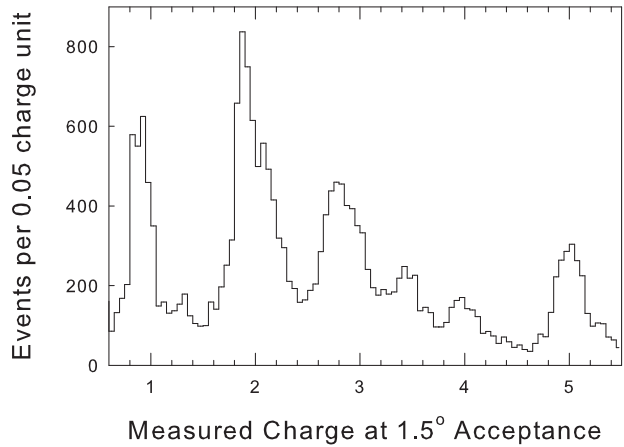}
\caption{\label{fig:fig4}Part of the charge spectrum from 1 GeV/nucleon $^{16}$O ions incident on a polyethylene
target of 2.82 g cm$^{-2}$ depth.
The H and He fragments have, on average, higher velocities than the primaries used to set the charge scale,
and so these peaks appear below the corresponding integer values. Other peaks, or shoulders on larger peaks, are
due to non-leading fragments and appear at effective charges of $\sim$ 1.3 (two H fragments), 2.1 (He + H), etc.}
\end{figure}

\subsection{Non-Leading He Fragments}
Many of the same points made in the preceding subsection apply to non-leading He fragments, but with a few important
differences: (1) the He fragment angular distributions are more forward-focused than those of H fragments, and (2) they 
make larger contributions to effective charge than do H fragments. Accordingly, the presence of non-leading and/or multiple 
He fragments can often be inferred from typical small-acceptance spectra, as was shown in Figure 2 for a $^{14}$N beam 
and again, with some important differences, in Figure 4 for a $^{16}$O beam. As discussed previously, the peak around 
charge 3.45 is due to the detection of three He fragments in coincidence (although a contribution from Li + He cannot 
be ruled out). The peak in the charge 2.6 to 3.2 region is due to events in which pairs of He fragments are detected 
in coincidence, along with events in which there is a leading Li fragment. In contrast to the spectrum in Figure 2, 
here we do not see a clear separation of the $Z$ $\approx$ 2.8 peak from the Li peak, and there also appears to be a 
comparatively larger share of events near 2.8. As in Figure 2, the peak near 1.3 is due to the detection of two charge-1 
fragments in coincidence, and here a small peak is seen near charge 1.6 (in the low-end tail of the He distribution), 
which is likely due to detection of three charge-1 fragments in coincidence. Finally, the peak seen near effective 
charge of 2.1 is almost certainly due to coincidences of helium and charge-1 fragments (as is the similar peak in Figure 2).

We do not count non-leading fragments as contributing to the production cross sections of a given species. To properly
perform that analysis would require a more sophisticated experiment. However, we will partially take account of the
non-integer peaks in the following by making use of the events in the peaks due to coincidences of two or three helium 
fragments. 

\subsection{Lithium/Helium-Pair Cross Sections}
In the region around $Z_{\text{frag}}$ = 3, there are contributions from events with either a leading lithium fragment or a 
coincidence of two helium fragments. Cross sections have been obtained by applying the average of corrections for fragment 
masses of 6 and 7 amu. Again, there is ambiguity here; the``charge 3'' peaks in all likelihood contain significant contributions
from $^8$Be (leading to a pair of $^4$He fragments), $^7$Li, $^6$Li, and pairs of $^4$He fragments that are produced independently.
Choosing to average $A$ = 6 and $A$ = 7 represents a best guess as to the midpoint of the acceptance for these events. 

\subsection{Three-Helium Fragment Production Cross Sections}
Although acceptance corrections are applied in all other cases, we choose not to apply them to the cross sections corresponding 
to the $Z$ $\approx$ 3.5 peaks. This is because we have no \textit{a priori} knowledge of the angular distributions of the fragments
observed in these events. It is possible to treat the three fragments as having been produced independently, i.e., with no mutual
correlations, in which case the corrections tend to be large. However, it may be that a significant fraction of these events 
arises from production of $^8$Be in conjunction with another helium fragment; the decay of $^8$Be produces two $^4$He fragments 
that together have a more forward-focused angular distribution than a single He fragment. Because there is negligible $Q$ in the 
$^8$Be decay, the initial forward-going trajectory of the fragment tends to be preserved by the two $^4$He decay products. The
probability for detection of all three He fragments also depends on the Coulomb multiple scattering the fragments undergo, which
occurs independently. The detection efficiency for these events is therefore a strong function of the production mechanism, the 
beam ion and energy, target material and depth, and the acceptance angle of the small-acceptance detectors used in the analysis. 
Of course, cross sections are by definition independent of the detection efficiencies. However in this instance, uncertainties 
about the production mechanism make it impossible to estimate angular acceptances with confidence, and we therefore present 
these cross sections without acceptance corrections. We will return to this subject repeatedly in the following as results from 
each beam ion are presented and discussed. In the subsequent discussion, it will be convenient to define two hypotheses: (1) the
three $^4$He fragments produced independently, so that the detection efficiency, $\varepsilon$, is simply the cube of 
the efficiency for detection 
of a single $^4$He fragment, i.e., $\varepsilon_{3He}$=$[\varepsilon$($^4$He)]$^3$; (2) two of the three helium fragments are 
the products of $^8$Be decay, and the third is produced independently, so that $\varepsilon_{3He}$=$[\varepsilon$($^4$He)]
$\times$ $[\varepsilon$($^8$Be)]. We will refer to these as hypothesis 1 and hypothesis 2.

\subsection{Beryllium Production Cross Sections}
The charge 4 category also presents some ambiguities. For the $^{14}$N beam, there is no plausible background from the
detection of four helium fragments in coincidence ($Z_{\text{eff}}$ = 16). However, for all other beams reported on here, such
events are at least theoretically possible, and if any such events occur they are indistinguishable from events in which
a single Be fragment is detected. For the $^{16}$O beam ions, fission into a pair of $^8$Be fragments, and subsequent
decay of the $^8$Be seems to be a plausible source of background. In the other direction, the detected Be cross section
cannot include any direct contribution from the production of a single $^8$Be fragment. When comparing to model calculations,
it is necessary to subtract the predicted $^8$Be contribution to the total Be production cross section. 

\section{Light Fragment Cross Sections}
In the following, we present the cross sections obtained with the ``small'' acceptances shown in Table I above. Since the
acceptances vary with beam ion, energy, and the angle subtended by the small acceptance detector in a given experiment,
in each subsection we show a plot of the detection efficiency vs. fragment mass number for each of the experiments
discussed therein. When applying corrections, we assume charge 4 fragments have mass 8, charge 2 fragments have mass 4,
and charge 1 fragments have mass 1. For charge 3, we average the results obtained for masses 6 and 7. The choice of
mass 8 for charge 4 represents a rough average of the acceptances for the stable isotopes with masses 7 and 9. The
results are somewhat sensitive to these choices, but in the absence of isotopic resolution in the experiments, we are 
guided by the NUCFRG2 model, since (unlike LAQGSM and PHITS) the code directly outputs isotopic cross sections. 

\subsection{Nitrogen Beams}
Table X shows the results for the production of Be and lighter fragments with the $^{14}$N beams. The results for
Be fragments were shown in Table VI above but are repeated here as they help to illuminate the overall trends.
As shown in Table I, the 290 MeV/nucleon data were obtained with an acceptance angle of 1.7$^\circ$ and the 400
MeV/nucleon data at 3.8$^\circ$. Figure 5 shows the calculated acceptances as a function of fragment mass number 
for the two nitrogen beams in the small-acceptance detectors. The larger efficiency in the 400 MeV/nucleon experiment 
is due both to the higher beam energy and to the larger acceptance angle. The acceptance corrections, applied to both 
data sets (except for the three-helium coincidence results) bring the charge 4 results into reasonable agreement at the 
two energies; the small differences seen for charge 4 could well be real, that is, due to actual weak energy dependence 
of the cross sections, although the differences are for the most part within or barely beyond one-sigma significance. 

\begin{figure}
\includegraphics[width=3.37in]{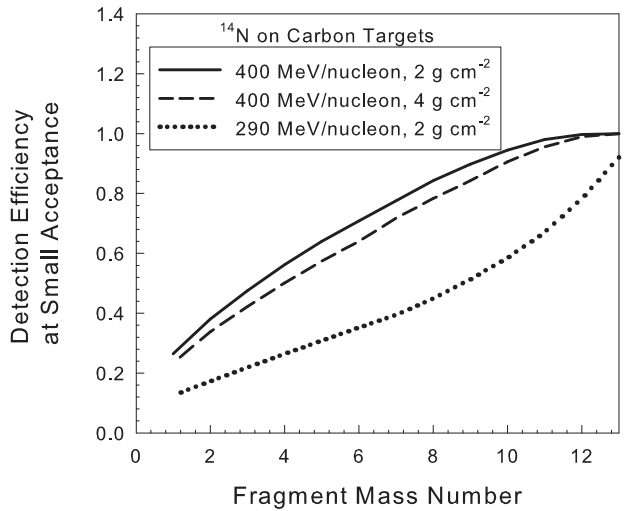}
\caption{\label{fig:fig5}Calculated acceptances for fragments in the small acceptance detectors for the 290 and
400 MeV/nucleon $^{14}$N experiments.}
\end{figure}

The uncorrected results for coincidences of three helium fragments are, in contrast, very different from one another - 
the 400 MeV/nucleon cross sections are larger by a factor of about 3 for all targets. This is due to the combined effects 
of the larger acceptance angle and the greater forward focusing of fragments produced by the higher-energy beam. As
described above, there is ambiguity about the production mechanism for these events, and it is not clear how the
cross sections should be corrected.  We find, suggestively, that in our acceptance model for three helium fragments, 
assuming no energy-dependence of the cross sections, the ratio of efficiencies predicted by hypothesis (1) for
these beams is a little over 5, and by hypothesis (2), exactly 3.0 \footnote{The calculation in this instance
was performed for a copper target, since copper is the closest thing to an ``average'' target (in terms of its Z and A) 
in these experiments. The results are only weakly dependent on the choice of target, and in other calculations
we used carbon as the target material.}. 

For the Li/He-pair category, the results for the two energies are mutually consistent within the uncertainties. 
There is no particular reason to expect this. Consider that the two-helium contribution to the peak likely consists 
of events in which there were actually three He fragments produced, but only two were detected, so there is ``feed-down'' 
from the three-helium category. There is also feed-down to the next category (one leading He fragment detected) - events 
in which two He fragments are produced, but only one is detected. Given this complexity, and the differences in acceptances
in the two experiments, it is surprising (and perhaps not meaningful) to find agreement.

For the single-helium category, the acceptance-corrected cross sections at 290 MeV/nucleon are all larger than those
at 400 MeV/nucleon. The ratios (400 MeV/nucleon cross sections divided by 290 MeV/nucleon cross sections) are all
mutually consistent, in the range 0.70 - 0.80, for carbon and heavier targets. The fact that cross sections in this
category appear larger at 290 MeV/nucleon is not surprising. At the lower energy, there is more feed-down from the two- 
and three-helium categories due to more fragments being outside the acceptance, and the acceptance corrections are 
substantially larger. Similar arguments apply to the $Z$ = 1 category, where again the cross sections obtained at
290 MeV/nucleon are larger than those at 400 MeV/nucleon. For carbon and heavier targets, ratios are again mutually
consistent, in the 0.78 to 0.93 range.

\begin{table*}
\caption{\label{table:tab10}Light fragment production cross section for $^{14}$N beams on elemental targets. Beam energies
at extraction are shown, in units of MeV/nucleon. All cross sections are in millibarns. The results for three helium fragments
detected in coincidence are not corrected for acceptance.}
\begin{ruledtabular}
\begin{tabular}
{c c c c c c c c}
$Z_{\text{frag}}$ & $E_{\text{beam}}$ & H          & C           & Al          & Cu           & Sn           & Pb \\
\hline
4          & 290        & 14 $\pm$ 2  & 48 $\pm$ 2 &  67 $\pm$ 3  &  97 $\pm$ 5  & 115 $\pm$ 9  & 133 $\pm$ 11 \\
4          & 400        & 15 $\pm$ 2  & 43 $\pm$ 2 &  59 $\pm$ 4  &  81 $\pm$ 5  & ~92 $\pm$ 7  & 130 $\pm$ ~9  \\
\hline
3 He coin. & 290        & 8 $\pm$ 1   & 18  $\pm$ 1 & 25  $\pm$ 2 & 28 $\pm$ 2   & 31  $\pm$ 4  & 28 $\pm$ 4\\
3 He coin. & 400        & 27 $\pm$ 3  & 54  $\pm$ 3 & 69  $\pm$ 4 & 80 $\pm$ 5   & 91  $\pm$ 7  & 93 $\pm$ 7 \\
\hline
3 or 2 He coin. & 290   & 49 $\pm$ 11 &  138 $\pm$ 9& 186 $\pm$ 13& 296 $\pm$ 25 & 319 $\pm$ 33 & 436 $\pm$ 46 \\
3 or 2 He coin. & 400   & 48 $\pm$ ~7 &  147 $\pm$ 7& 177 $\pm$ 10& 268 $\pm$ 14 & 330 $\pm$ 20 & 397 $\pm$ 23 \\
\hline
2          & 290        & 97 $\pm$ 26& 337 $\pm$ 21& 482 $\pm$ 32 & 776 $\pm$ 63 & 841 $\pm$ 83 & 1222 $\pm$ 123 \\
2          & 400        & 53 $\pm$ 11& 267 $\pm$ 13& 345 $\pm$ 19 & 543 $\pm$ 26 & 673 $\pm$ 37 & ~927 $\pm$ 48 \\
\hline
1          & 290        & 92 $\pm$ 29& 399 $\pm$ 26& 712 $\pm$ 48 & 1194$\pm$ 98 & 1549 $\pm$ 153 & 2330 $\pm$ 232 \\
1          & 400        & 26 $\pm$ 15& 401 $\pm$ 19& 569 $\pm$ 31 & ~915$\pm$ 45 & 1211 $\pm$ ~66 & 1839 $\pm$ ~81 
\label{tab10}
\end{tabular}
\end{ruledtabular}
\end{table*}

\subsection{Oxygen Beams}
Table XI shows the results for Be and lighter fragments with $^{16}$O beams at four energies. The
corrected Be cross sections show no consistent pattern of energy dependence. For Sn targets, there
is an increase with increasing beam energy, but this is not seen for other targets.

To understand the cross sections in the other categories, it is useful to note that for the 290, 600, 
and 1000 MeV/nucleon experiments, the small-acceptance detector angles were comparable, 1.7$^\circ$, 
1.4$^\circ$, and 1.4$^\circ$, respectively, while for the 400 MeV/nucleon experiment, the small acceptance 
detector subtended a half-angle of 2.5$^\circ$. The calculated acceptances are shown in Figure 6.
The curves for the 290 and 600 MeV/nucleon experiments sit almost on top of one another; this is
fortuitous, and not by design.

\begin{figure}
\includegraphics[width=3.37in]{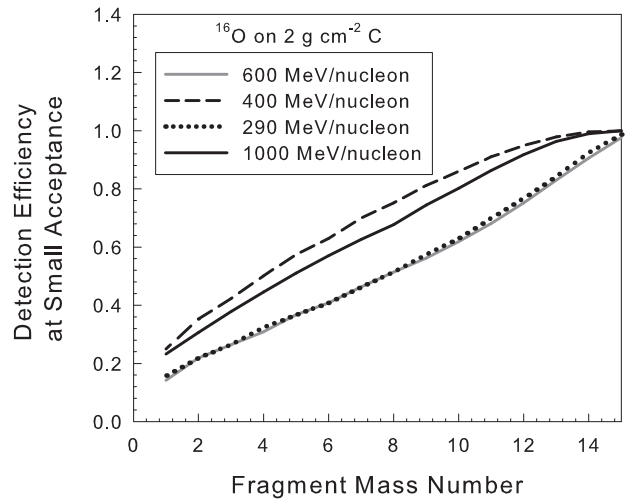}
\caption{\label{fig:fig6}Calculated acceptances for fragments in the small acceptance detectors for the four 
$^{16}$O experiments.}
\end{figure}

\begin{table*}
\caption{\label{table:tab11}Light fragment production cross section for $^{16}$O beams on elemental targets. Beam energies
at extraction are shown, in units of MeV/nucleon. All cross sections are in millibarns. The results for three helium fragments
detected in coincidence are not corrected for acceptance.}
\begin{ruledtabular}
\begin{tabular}
{c c c c c c c c}
$Z_{\text{frag}}$ & $E_{\text{beam}}$ & H          & C           & Al          & Cu           & Sn           & Pb \\
\hline
4          & 290        & ~8 $\pm$ 2  & 49 $\pm$ 3 &  66 $\pm$ 4  & 71 $\pm$ 5  & ~68 $\pm$ 8  & 114 $\pm$ 16 \\
4          & 400        & 16 $\pm$ 3  & 46 $\pm$ 4 &  61 $\pm$ 4  & 75 $\pm$ 6  & ~87 $\pm$ 8  & ~98 $\pm$ ~9  \\
4          & 600        & 14 $\pm$ 3  & 45 $\pm$ 4 &  53 $\pm$ 3  & 66 $\pm$ 5  & ~96 $\pm$ 8  & 112 $\pm$ 11  \\
4          & 1000       & 16 $\pm$ 3  & 50 $\pm$ 4 &  54 $\pm$ 3  & 76 $\pm$ 4  & 109 $\pm$ 7  & 117 $\pm$ ~9  \\
\hline
3 He coin. & 290        & ~5 $\pm$ 1  & 17  $\pm$ 1 & 20  $\pm$ 1 & 17 $\pm$ 2   & 26  $\pm$ 3  & 22 $\pm$ 4\\
3 He coin. & 400        & 15 $\pm$ 3  & 47  $\pm$ 4 & 56  $\pm$ 4 & 61 $\pm$ 5   & 64  $\pm$ 6  & 55 $\pm$ 5 \\
3 He coin. & 600        & ~7 $\pm$ 1  & 20  $\pm$ 2 & 20  $\pm$ 1 & 23 $\pm$ 2   & 26  $\pm$ 3  & 29 $\pm$ 3 \\
3 He coin. & 1000       & 15 $\pm$ 2  & 33  $\pm$ 2 & 41  $\pm$ 2 & 48 $\pm$ 3   & 48  $\pm$ 3  & 59 $\pm$ 5 \\
\hline
3 or 2 He coin. & 290   & 30 $\pm$ 6 &  125 $\pm$ ~7& 143 $\pm$ 11& 193 $\pm$ 15 & 210 $\pm$ 20 & 271 $\pm$ 28 \\
3 or 2 He coin. & 400   & 36 $\pm$ 9 &  136 $\pm$ 10& 168 $\pm$ 10& 205 $\pm$ 16 & 240 $\pm$ 19 & 308 $\pm$ 23 \\
3 or 2 He coin. & 600   & 33 $\pm$ 6 &  107 $\pm$ ~8& 118 $\pm$ ~6& 161 $\pm$ ~9 & 209 $\pm$ 13 & 225 $\pm$ 17 \\
3 or 2 He coin. & 1000  & 46 $\pm$ 7 &  129 $\pm$ ~9& 160 $\pm$ ~8& 220 $\pm$ 11 & 267 $\pm$ 15 & 294 $\pm$ 19 \\
\hline
2          & 290        & ~67 $\pm$ 15& 357 $\pm$ 18& 474 $\pm$ 34 & 608 $\pm$ 43 & 702 $\pm$ 60 & ~904 $\pm$ 92 \\
2          & 400        & ~45 $\pm$ 15& 270 $\pm$ 19& 335 $\pm$ 20 & 461 $\pm$ 33 & 582 $\pm$ 43 & ~820 $\pm$ 56 \\
2          & 600        & 103 $\pm$ 20& 410 $\pm$ 29& 497 $\pm$ 24 & 592 $\pm$ 31 & 961 $\pm$ 54 & 1224 $\pm$ 78 \\
2          & 1000       & ~70 $\pm$ 14& 278 $\pm$ 19& 377 $\pm$ 17 & 501 $\pm$ 25 & 695 $\pm$ 37 & ~819 $\pm$ 49 \\
\hline
1          & 290        & 34 $\pm$ 16& 385 $\pm$ 20 & 607 $\pm$ 44 &  781$\pm$ 57 & 1033 $\pm$ ~90 & 1426 $\pm$ 148 \\
1          & 400        & 21 $\pm$ 17& 325 $\pm$ 23 & 525 $\pm$ 31 &  806$\pm$ 57 & ~970 $\pm$ ~71 & 1471 $\pm$ 101 \\ 
1          & 600        & 97 $\pm$ 25& 515 $\pm$ 38 & 719 $\pm$ 35 &  967$\pm$ 52 & 1487 $\pm$ ~86 & 2017 $\pm$ 128 \\
1          & 1000       & 54 $\pm$ 18& 404 $\pm$ 28 & 576 $\pm$ 27 &  836$\pm$ 42 & 1274 $\pm$ ~68 & 1585 $\pm$ ~94 
\label{tab11}
\end{tabular}
\end{ruledtabular}
\end{table*}

The effects of the acceptance differences can be seen in the uncorrected cross sections for three helium 
fragments. Results with the 290 and 600 MeV/nucleon beams are mutually consistent, which makes sense 
in view of the nearly-identical acceptance curves for the two experiments, with the additional assumption 
that the production cross section is weakly or not at all dependent on beam energy. For the 600 and 1000 MeV/nucleon 
data, where the acceptance angle was the same, the higher-energy beam yields larger cross sections in all cases, 
typically by a factor close to 2, due to the greater forward boost, which again is reflected in the acceptance
curve in Figure 6. Compared to results at other energies, the 400 MeV/nucleon cross sections are significantly 
larger than those obtained at 290 and 600 MeV/nucleon, by roughly a factor of 2, and in most cases are larger 
than those obtained at 1000 MeV/nucleon. This can only be due to the comparatively large acceptance angle 
employed in the 400 MeV/nucleon experiment; the curves in Figure 6 bear this out. The acceptance model
predicts that, for hypothesis 1, the cross section measured in the 290 MeV/nucleon experiment should be only 
25\% as large as that measured in the 400 MeV/nucleon experiment. For hypothesis 2, $^8$Be + $^4$He, the same 
ratio is predicted to be 43{\%}, somewhat closer to (but slightly higher than) the observed ratios. These ratios 
suggest that the observed events are a mix of the two types, with a majority of $^8$Be events. Comparing the 
600 and 1000 MeV/nucleon results to those obtained at 400 MeV/nucleon yields a muddled picture: the 600 MeV/nucleon 
data are also consistent with a mix of the event types, but the 1000 MeV/nucleon data are consistent with being 
entirely due to $^8$Be + $^4$He. 

Turning to the Li or 2 He cross sections, we find a high degree of consistency between the 400 and 1000
MeV/nucleon experiments. At first glance, this is slightly surprising since the acceptances in the 400 
MeV/nucleon experiment are larger by 10-20{\%}, as can be seen in Figure 6. The cross sections obtained
with the 290 MeV/nucleon beam are marginally consistent with, but in all cases smaller than, those obtained
at 400 and 1000 MeV/nucleon, and the 600 MeV/nucleon cross sections are (for the most part) smaller still.

Based on the acceptance curves, we might expect the 600 MeV/nucleon cross sections to be consistent in
all cases with those from the 290 MeV/nucleon beam, but this is not what is observed. For the single-He category, 
the cross sections obtained at 600 MeV/nucleon are generally the largest. Figure 7 shows the cross sections as 
functions of target mass for the (leading) $Z$ = 2 and $Z$ = 1 categories. For both, the 600 MeV/nucleon 
cross sections are larger than those at the other energies, particularly for the Sn and Pb targets. In view of 
the similarity in the detection efficiencies between this experiment and the 290 MeV/nucleon experiment, this 
may seem surprising, but referring to Tables II and VII is illuminating. The charge-changing cross sections tend 
to be slightly larger at 600 MeV/nucleon compared to 290 MeV/nucleon (Table II), and the cross sections in the
$Z~\leq$ 5 category are 10-20\% larger at the higher energy (Table VII). Small differences in the production 
cross sections are multiplied by large acceptance corrections to produce the large differences seen in Figure 7.
And although the grouped $Z$$\leq$ 5 production cross sections are found to be approximately equal at 600 and 1000 
MeV/nucleon (Table VII), the higher-energy results receive significantly smaller corrections.

\begin{figure*}
\includegraphics[width=6.0in]{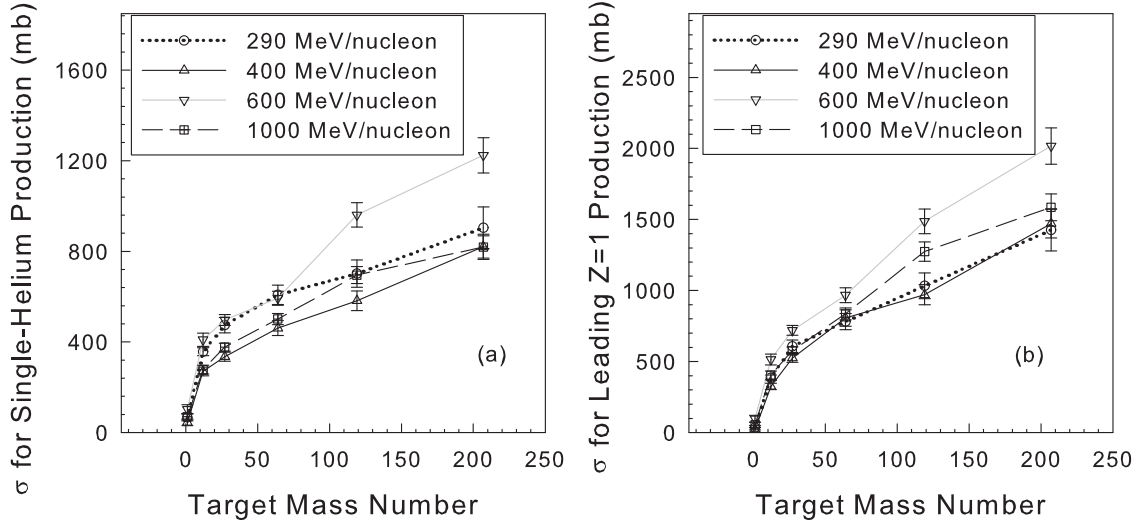}
\caption{\label{fig:fig7} Corrected cross sections for events with a single leading charge 2 fragment (left)
and a leading charge 1 fragment (right) as measured at small acceptance detectors for the four $^{16}$O experiments.}
\end{figure*}

\subsection{Neon Beams}
In Table XII, we show results for charge 4 and lighter fragments with $^{20}$Ne beams at three energies. 
Acceptance curves for the three beam energies are shown, for a 2 g cm$^{-2}$ $^{12}$C target, in Figure 8. 
There were two experiments with the 600 MeV/nucleon beam, one of which was reported on earlier \cite{Zeitlin2001}.
Small-acceptance cross sections shown here are the result of combining the data sets as described above, except
for the three helium-fragment coincidence category, for which we show separately results obtained with acceptances 
of 2.5$^\circ$ and 1.7$^\circ$. 

The charge 4 production cross sections increase with increasing energy for the hydrogen target, but for other targets
tend to decrease slightly with increasing energy. This is quite similar to the behavior seen for smaller charge changes
in Tables VII and VIII. The cross sections for three helium fragments detected in coincidence all increase with energy,
but this may be largely an acceptance effect. Since we do not know \textit{a priori} whether or not there is any
dependence of the underlying production cross section on beam energy, the fact that we have measurements at two
acceptance angles using the 600 MeV/nucleon beam is potentially instructive. For all targets, the cross sections 
for 600 MeV/nucleon beams are found to be larger when the acceptance angle is 2.5$^\circ$ compared to 1.7$^\circ$,
which is as expected. Taking the ratio of the cross sections at 2.5$^\circ$ to those at 1.7$^\circ$ by target,
we see values ranging from 1.2 to 2.2, but on closer inspection we find the data are reasonably consistent with 
being independent of the target. The weighted average of the ratios is 1.57 $\pm$ 0.09, with a $\chi^2$ of
7.3 for 5 degrees of freedom for the ratio being target-independent. Our acceptance model predicts, using the 
same data points that populate Figure 8, a ratio of 1.74 for hypothesis 2 and a ratio of 2.82 for hypothesis 1. 
The data are clearly - again - much more consistent with hypothesis 2 ($^8$Be + $^4$He). At 290 and 400 MeV/nucleon, 
the cross sections in this category are (except for the hydrogen target) significantly larger at the higher energy. 
This is due to the larger acceptance in the 400 MeV/nucleon experiment illustrated in Figure 8.

For the Li or He pair category, there is a consistent ordering of the cross sections, with those obtained at
290 MeV/nucleon being the largest for all targets and those obtained at 600 MeV/nucleon being the smallest. 
The 400 MeV/nucleon data are in between, but are mostly close to the 600 MeV/nucleon results. Referring to
Table VIII, the (uncorrected) cross sections for the grouped $Z$$\leq$ 5 category are seen to be mutually 
consistent within uncertainties, suggesting that the ordering of the cross sections in this category is
likely due to the acceptance corrections and not to any energy dependence in the underlying production 
mechanism. Similar remarks apply to the $Z_{\text{frag}}$ = 2 and 1 categories, although in the latter we are
unable to report cross sections from the 290 MeV/nucleon experiment. 

\begin{figure}
\includegraphics[width=3.37in]{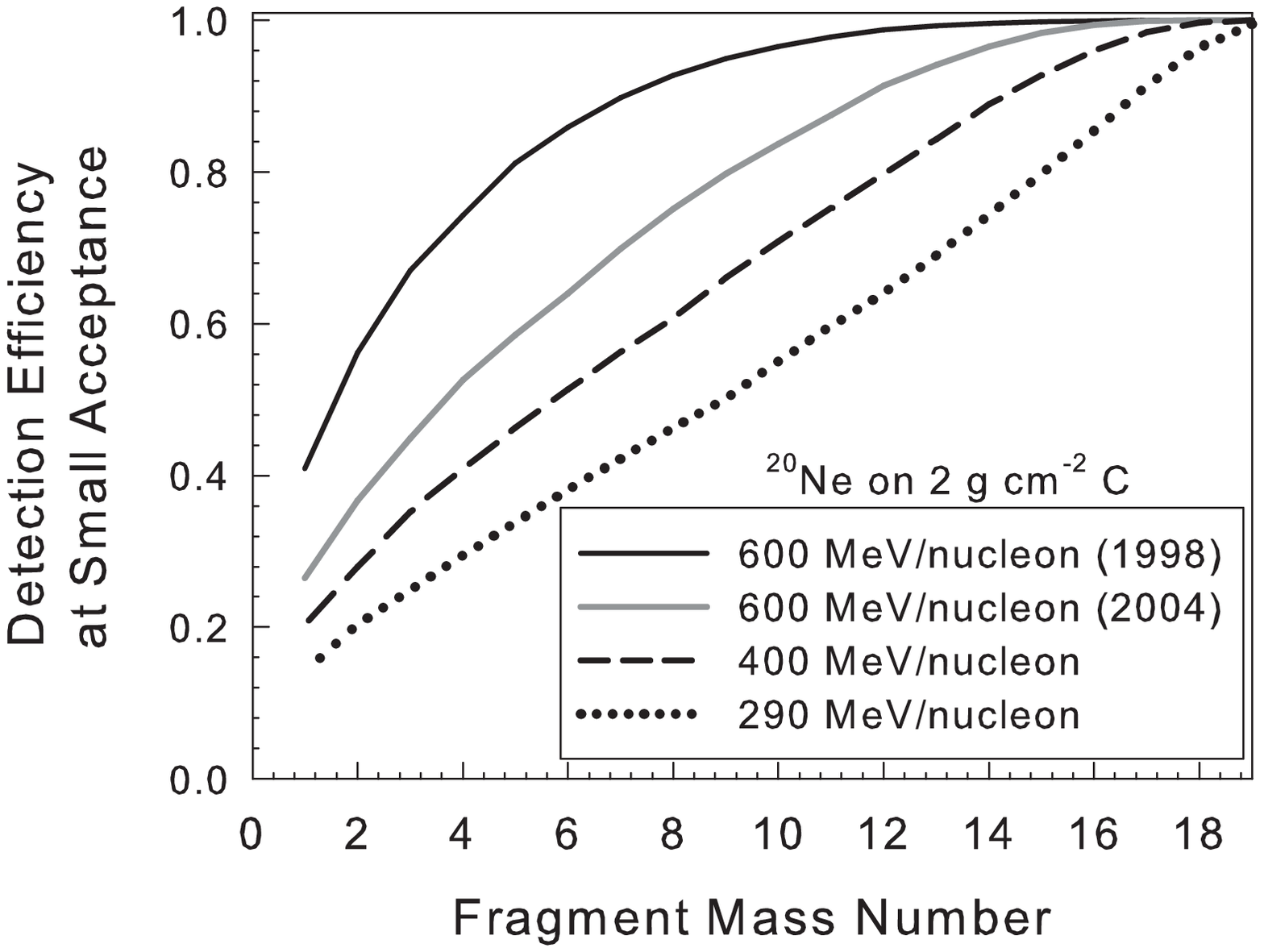}
\caption{\label{fig:fig8}Acceptance vs. fragment mass number as calculated for the small acceptance detectors 
for the four $^{20}$Ne experiments.}
\end{figure}

\begin{table*}
\caption{\label{table:tab12}Light fragment production cross section for $^{20}$Ne beams on elemental targets. Beam energies
at extraction are shown, in units of MeV/nucleon. All cross sections are in millibarns. The results for three helium fragments
detected in coincidence are not corrected for acceptance. For the 290 MeV/nucleon beam, fragments with charge 1 could not
be resolved.}
\begin{ruledtabular}
\begin{tabular}
{c c c c c c c c}
$Z_{\text{frag}}$ & $E_{\text{beam}}$ & H          & C           & Al          & Cu           & Sn           & Pb \\
\hline
4          & 290        & ~6 $\pm$ 1  & 52 $\pm$ 4 &  66 $\pm$ 5  & 96 $\pm$ 8  & 130 $\pm$ 14 & 135 $\pm$ 17 \\
4          & 400        & ~9 $\pm$ 2  & 46 $\pm$ 3 &  59 $\pm$ 3  & 70 $\pm$ 4  & ~82 $\pm$ ~7 & 103 $\pm$ ~9  \\
4          & 600        & 14 $\pm$ 2  & 46 $\pm$ 3 &  59 $\pm$ 4  & 68 $\pm$ 5  & 102 $\pm$ ~7 & ~96 $\pm$ ~7  \\
\hline
3 He coin. & 290        & ~8 $\pm$ 1  & 14  $\pm$ 2 & 21  $\pm$ 1 & 13 $\pm$ 2   & ~16  $\pm$ ~3  & 14 $\pm$ 4\\
3 He coin. & 400        & ~8 $\pm$ 1  & 21  $\pm$ 1 & 25  $\pm$ 1 & 31 $\pm$ 1   & ~35  $\pm$ ~3  & 30 $\pm$ 3 \\
3 He coin. & 600  (2.5$^\circ$)  & 16 $\pm$ 2  & 56  $\pm$ 1 & 58  $\pm$ 2 & 70 $\pm$ 5   & 112  $\pm$ ~6  & 81 $\pm$ 4 \\
3 He coin. & 600  (1.7$^\circ$)  & 14 $\pm$ 3  & 31  $\pm$ 1 & 37  $\pm$ 2 & 53 $\pm$ 3   & ~51  $\pm$ ~5  & 55 $\pm$ 5 \\
\hline
3 or 2 He coin. & 290   & 27 $\pm$ 2 &  161 $\pm$ 10 & 257 $\pm$ 17& 371 $\pm$ 28 & 294 $\pm$ 28 & 632 $\pm$ 66 \\
3 or 2 He coin. & 400   & 20 $\pm$ 6 &  119 $\pm$ ~6 & 149 $\pm$ ~7& 191 $\pm$ 10 & 229 $\pm$ 15 & 272 $\pm$ 18 \\
3 or 2 He coin. & 600   & 19 $\pm$ 7 &  103 $\pm$ ~8 & 128 $\pm$ 10& 163 $\pm$ 14 & 219 $\pm$ 19 & 262 $\pm$ 22 \\
\hline
2          & 290        &  73 $\pm$ 4 & 361 $\pm$ 22& 526 $\pm$ 33 & 973 $\pm$ 65 & 1293 $\pm$ 106 & 1632 $\pm$ 150 \\
2          & 400        &  40 $\pm$ 14& 298 $\pm$ 15& 402 $\pm$ 19 & 547 $\pm$ 28 & ~604 $\pm$ ~37 & ~881 $\pm$ ~53 \\
2          & 600        &  18 $\pm$ 11& 190 $\pm$ 15& 265 $\pm$ 21 & 378 $\pm$ 32 & ~434 $\pm$ ~37 & ~588 $\pm$ ~47 \\
\hline
1          & 400        & ~8 $\pm$ 18 & 392 $\pm$ 19& 572 $\pm$ 26 &  809$\pm$ 51 & ~998 $\pm$ 65  & 1476 $\pm$ 96 \\ 
1          & 600        & 11 $\pm$ 14 & 247 $\pm$ 19& 455 $\pm$ 36 &  680$\pm$ 57 & ~871 $\pm$ 72  & 1232 $\pm$ 99 
\label{tab12}
\end{tabular}
\end{ruledtabular}
\end{table*}

\subsection{400 MeV/nucleon $^{24}$Mg}
Table XIII shows the cross sections obtained for the 400 MeV/nucleon $^{24}$Mg beam, and Figure 9 shows
the results of the acceptance calculation for this beam on a 4 g cm$^{-2}$ $^{12}$C target as was used
in the experiment. Also shown in Figure 9 are the efficiency curves found for the other experiments with
400 MeV/nucleon beams reported here. The inset figure zooms in on the fragment charge range from 1 to 4,
which we examine in the following. Again, the acceptances are only weakly dependent on the type and depth 
of the target material, so the curves in Figure 9 are representative of all runs with these beams. Note that 
if all other variables are held constant except for the mass of the beam ion, as that mass increases the 
acceptance of the lightest fragments decreases, because the width of the angular distribution is driven by 
a term proportional to ($A_{\text{beam}} - A_{\text{frag}}$)$^{1/2}$. 

Since data are only available with a single beam energy for $^{24}$Mg, we cannot probe these results
for possible energy dependence of the production cross sections. However, other features of the data
merit comment. For the hydrogen target, the production cross sections for both $Z_{\text{frag}}$ = 1 and 
$Z_{\text{frag}}$ = 2 are consistent with 0. This is seen for $Z_{\text{frag}}$ = 1 for several of the other beams 
reported above, but not for $Z_{\text{frag}}$ = 2. The large uncertainties in these data do not 
permit us to make a stronger statement; comparing to the 400 MeV/nucleon $^{20}$Ne result for $Z_{\text{frag}}$ = 2, 
we can only say that the cross section with the $^{24}$Mg beam is smaller at about the 1.5$\sigma$ level.
It would be useful to obtain more data with this beam ion, at 400 MeV/nucleon and other energies,
particularly since Mg is prominent among the GCR heavy ions. 

Comparing these data to the 400 MeV/nucleon $^{20}$Ne cross sections, we find very similar values
for the $Z_{\text{frag}}$ = 4 cross sections for all targets, and also for the three-helium-coincidence category.
For the lower-$Z$ categories, excluding the hydrogen target data, the cross sections for the $^{24}$Mg 
beam are significantly larger than for the $^{20}$Ne beam, and also for the other beam ions at 400
MeV/nucleon. The same can also be said for the $Z_{\text{frag}}$ = 1 category. Figures 10 and 11 illustrate 
these trends. Cross sections for the beams other than Mg are in many instances mutually consistent,
with the $^{14}$N beam data tending to be slightly higher than the $^{16}$O and $^{20}$Ne data for
$Z_{\text{frag}}$ = 1. The $^{24}$Mg cross sections are larger than those for other beams by factors of
about 1.5 for $Z_{\text{frag}}$ = 2 and factors of about 2 for $Z_{\text{frag}}$ = 1. Referring to the inset figure
in Fig. 9, we see that the detection efficiencies for these light fragments were lowest in the Mg
experiment. We expect that production cross sections for protons and helium must increase with increasing 
beam charge and mass since there are simply more nucleons available to be sheared off the projectile, 
and the probability for detecting one or more fragments should increase. 
 
\begin{figure}
\includegraphics[width=3.37in]{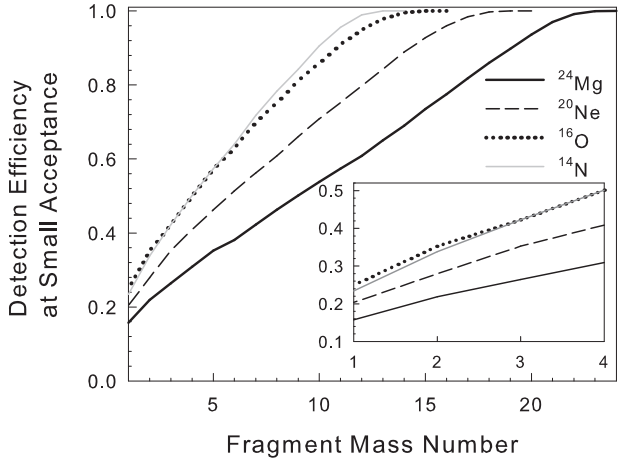}
\caption{\label{fig:fig9}Acceptance vs. fragment mass number as calculated for the small acceptance detectors 
for the experiments with 400 MeV/nucleon beam energies and carbon targets.}
\end{figure}

\begin{table*}
\caption{\label{table:tab13}Light fragment production cross section for a 400 MeV/nucleon $^{24}$Mg beam 
on elemental targets. All cross sections are in millibarns. The results for three helium fragments
detected in coincidence are not corrected for acceptance.}
\begin{ruledtabular}
\begin{tabular}
{c c c c c c c}
$Z_{\text{frag}}$      & H           & C             & Al             & Cu           & Sn           & Pb \\
\hline
4               & ~7 $\pm$ ~2 & ~41 $\pm$ ~3  & ~~55 $\pm$ ~4  & ~~88 $\pm$ ~~8 & ~~78 $\pm$ ~~9 & ~126 $\pm$ ~16 \\
\hline
3 He coin.      & ~6 $\pm$ ~1 & ~15 $\pm$ ~1  & ~~25 $\pm$ ~2  & ~~28 $\pm$ ~~3 & ~~28 $\pm$ ~~3 & ~~31 $\pm$ ~~5\\
\hline
3 or 2 He coin. & 17 $\pm$ ~7 & 139 $\pm$ ~9  & ~183 $\pm$ 13  & ~264 $\pm$ ~20 & ~274 $\pm$ ~25 & ~321 $\pm$ ~35 \\
\hline
2               & ~0 $\pm$ 19 & 428 $\pm$ 27  & ~586 $\pm$ 40  & ~783 $\pm$ ~57 & 1074 $\pm$ ~89 & 1470 $\pm$ 138 \\
\hline
1               & 25 $\pm$ 27 & 568 $\pm$ 37  & 1109 $\pm$ 75  & 1957 $\pm$ 140 & 2545 $\pm$ 209 & 4264 $\pm$ 389  
\label{tab13}
\end{tabular}
\end{ruledtabular}
\end{table*}

\begin{figure}
\includegraphics[width=3.37in]{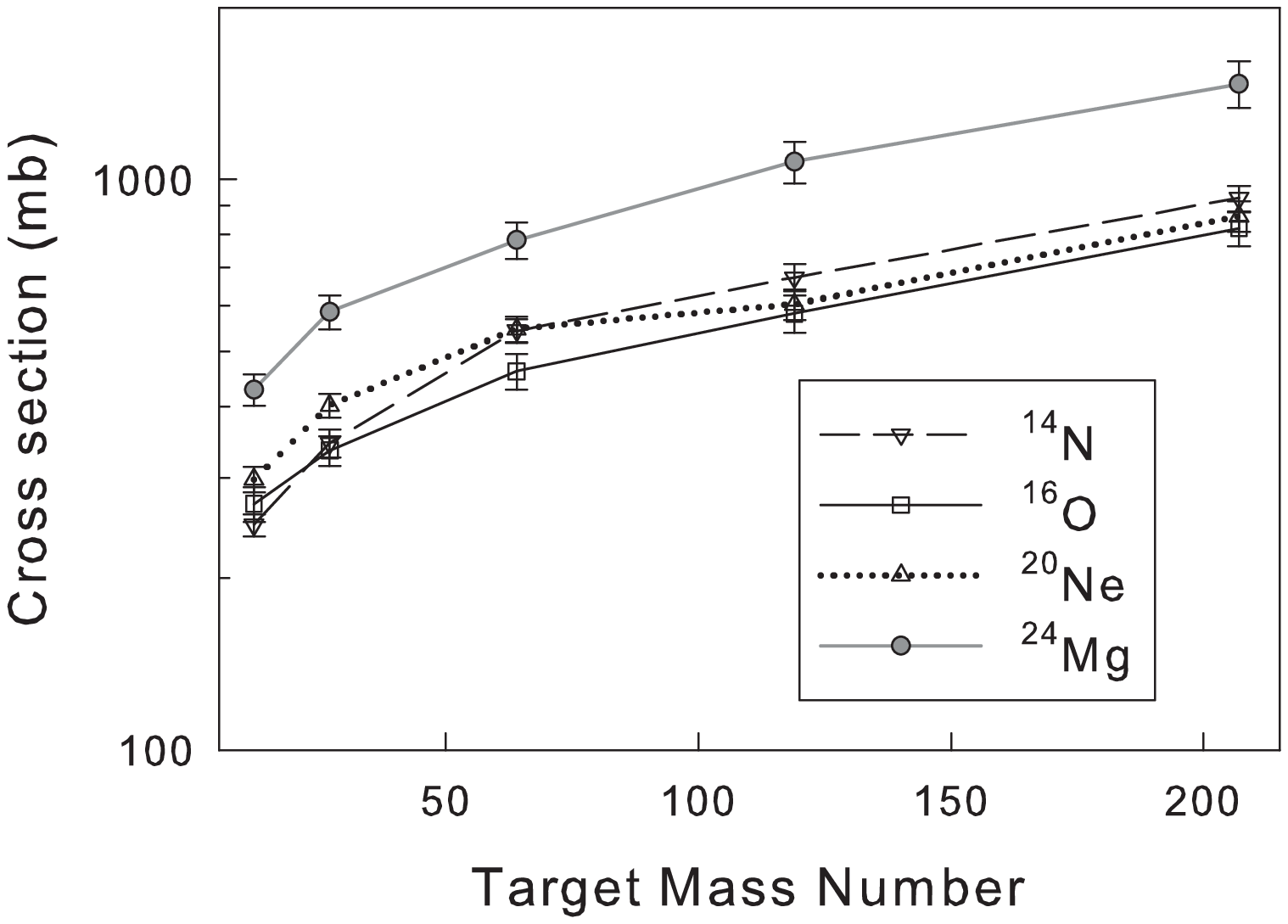}
\caption{\label{fig:fig10}Acceptance-corrected production cross sections for the leading helium category
for the 400 MeV/nucleon beams. Hydrogen-target data are excluded.}
\end{figure}

\begin{figure}
\includegraphics[width=3.37in]{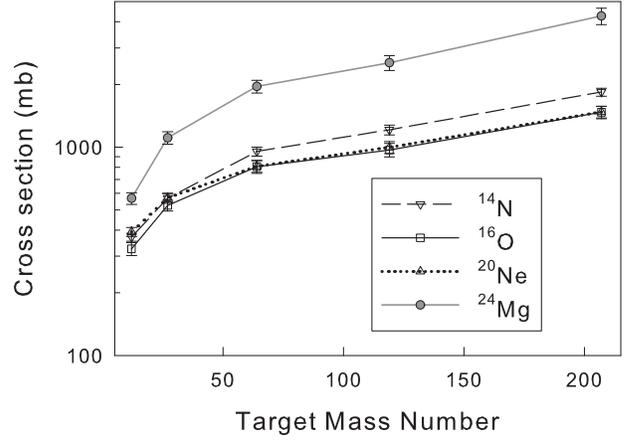}
\caption{\label{fig:fig11}Acceptance-corrected production cross sections for the leading $Z$ = 1 category
for the 400 MeV/nucleon beams. Hydrogen-target data are excluded.}
\end{figure}

\section{Fragment Cross Section Comparisons}
Articles in the literature for the beams studied here (and similar beams) have reported fragment cross
sections for small charge changes, limiting the scope of model comparisons. Here, we are able to extend
the comparisons down to $Z$ = 2, albeit with somewhat large uncertainties owing to the nature of the
measurements. We do not believe the $Z$ = 1 data reported above are suitable for model comparisons
because there is no feasible way to perform multiplicity weighting, mostly because a dominant share 
of the charge 1 fragments produced are undetectable since they accompany much heavier fragments.

\subsection{Model Descriptions of Fragmentation Processes}
Of the three models used here, PHITS and LAQGSM are three-dimensional Monte Carlo codes that simulate individual 
interactions in great detail. In contrast, NUCFRG2 is an engineering code designed for computational speed;
its physics content is based on an abrasion-ablation formulation with its free parameters tuned to reproduce
p-nucleus cross section data.

In LAQGSM, when the mass number of the excited nucleus is greater than 12, a three-stage process is modeled: 
intranuclear cascade (INC); preequilibrium emission of fragments from the excited remnant nucleus; and evaporation
and/or fission, if the compound nucleus is heavy enough to fission. 
When the mass number of the excited nucleus is 12 or less, LAQGSM uses the Fermi break-up model after INC. That is, 
the Fermi breakup model is used before the preequilibrium stage, as well as during the preequilibrium and evaporation 
stages, if the mass number of the excited nucleus becomes less than 13 due to emission of some preequilibrium and/or 
equilibrium particles. 

For the beam ions studied here, a significant share of the reactions
fall into the latter category and therefore test the Fermi break-up model. The latest version of the LAQGSM (03.03) 
incorporates an improved version of the Dubna Cascade Model \cite{Toneev1983}. It uses a continuous nuclear density 
distribution and experimental cross sections at energies below 4.5 GeV/nucleon. It has previously been shown \cite{Mashnik2008a, Mashnik2008b}
that there is good agreement between LAQGSM version 03.03 calculations and data taken by our group with $^{28}$Si beams at 
290, 600, and 1200 MeV/nucleon \cite{Zeitlin2007a} for fragment production cross sections. 

PHITS uses a cross section model developed by Tripathi to determine interaction probabilities, the Jaeri Quantum
Molecular Dynamics (JQMD) model to describe nucleus-nucleus collisions, and the Generalized Evaporation Model (GEM) 
to model fission and evaporation processes. Several previous comparisons between our data and PHITS can be found 
in the literature \cite{LaTessa2005, Sato2005, Mancusi2007, Zeitlin2007b, Zeitlin2008, Zeitlin2010}. 

\subsection{Production Cross Sections for Charges 2 and 3}
In the preceding, we reported cross sections in three categories that pertain to Li and He production: ``3 He'', 
``Li/2 He'', and ``$Z$ = 2.'' Cross sections obtained in the latter two categories are corrected for acceptance assuming 
fragment mass numbers of 7 and 4, respectively. Multiplicity-weighted estimates of the total He production cross sections
can be made with two opposing assumptions, in the hope of bracketing the true cross sections. For a given beam
ion/energy/target combination, our lower-bound estimate is obtained by multiplying the ``3 He'' cross section 
by three and adding it to the ``$Z$ = 2'' cross section. The upper-bound estimated is obtained by multiplying
the``Li/2 He'' cross section by two and adding this to the lower-bound estimate. These estimates do not include
contributions from non-leading He fragments that are undetectable with our experimental methods, so it is
conceivable that even our upper-bound estimates are too small. Also, there is no acceptance correction for
the ``3 He'' category, which would tend to cause an error in the same direction. An error in the other
direction stems from the implicit assumption that there is no lithium production whatsoever, which cannot
be true.

On average, the upper-bound estimates give better agreement with the models, particularly for PHITS and LAQGSM. 
That is, the agreement is not necessarily better with the upper bound estimate for each data point, but lower 
cumulative uncertainty (defined below) is obtained for all three models when the predictions are compared to the
upper-bound estimates.

For the $^{14}$N and $^{16}$O beams, the estimates of total He production appear to be largely energy-independent. 
Both upper- and lower-bound values for a given target tend to be mutually consistent across beam energies. 
Only the upper-bound values for hydrogen targets show a statistically-significant slope (a slight rise) when 
plotted as a function of the beam energy. The left-hand plot in Figure 12 shows the dependences of these estimates 
for $^{16}$O beams on H, C, and Al targets. Similar results are obtained for Cu, Sn, and Pb targets, but only the 
results for the three lighter targets are shown in order to keep the plot readable. In view of the varying 
experimental acceptances (see Figure 8 above), this is quite an interesting result. However, the $^{20}$Ne upper 
and lower bounds for total helium production do not quite show the same behavior. As can be seen in the plot on 
the right-hand side of Figure 12, the results with the 400 and 600 MeV/nucleon beams tend to be mutually consistent, 
but the cross section estimates with the 290 MeV/nucleon beam are in all cases significantly larger. 

\begin{figure*}
\includegraphics[width=5.7in]{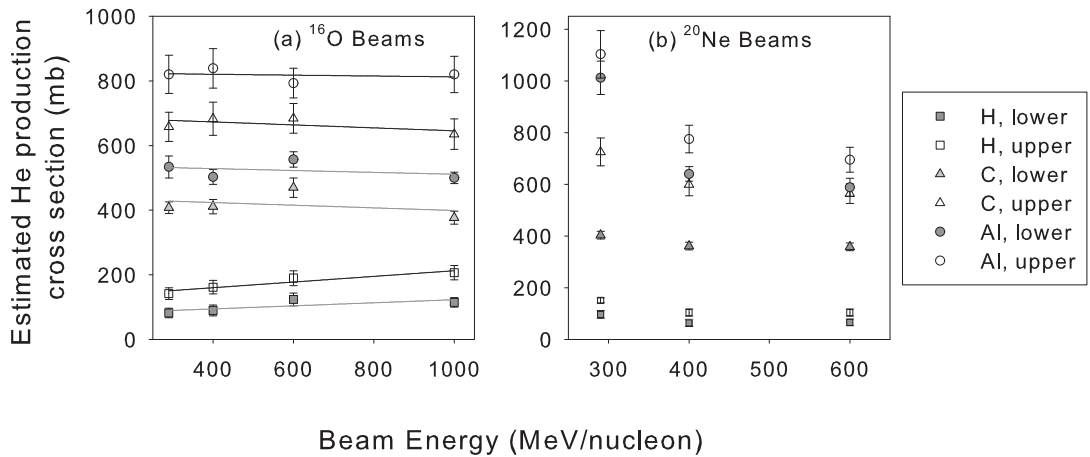}
\caption{\label{fig:fig12}Energy dependence of helium production cross sections as estimated by methods described 
in the text, H, C, and Al targets. Both upper- and lower-bound estimates from the data are shown. Left, in Fig. 12(a), 
results for $^{16}$O beams; right, (b), for $^{20}$Ne beams. In Fig. 12(a), the lines shown are for linear fits.}
\end{figure*}

To estimate the Li cross sections, we simply take half the measured cross section for the ``Li/2 He'' category
and assign a relative error of $\pm$ 33{\%}. This allows for the extreme (and implausible) possibilities that either
all of these events are Li, or that none of them is Li, at the three-sigma level. 

Model comparisons to the $Z$ = 1 data are excluded from consideration here for reasons mentioned above. Cross sections
for $Z$ = 1 predicted by PHITS and LAQGSM are typically factors of 2-3 larger than measured cross sections. For NUCFRG2,
the discrepancies are typically factors of 10-20, with the model predictions again being larger than the measured. 

\subsection{Model Predictions of Helium Production by $^{16}$O and $^{20}$Ne Beams}
Before proceeding to the overall comparisons of fragment cross sections and models, here we focus on the estimated 
helium production cross sections compared to the models. Given the difficulties involved in obtaining the estimates, 
the relatively large uncertainties associated with them, and the fact that we do not include non-leading He fragments
in our measurement, it is reasonable to wonder whether there is any correspondence at all with the model predictions. 

The cross sections in Tables X through XII that involve helium fragments are somewhat difficult to interpret, largely 
because the varying acceptances in the different experiments lead to disparate results for the different categories. 
However, the method of bounding the production cross sections described above appears to at least partially compensate 
for the acceptance effects, based on the results shown in Figure 12 and on the reasonable agreement found with PHITS 
and LAQGSM predictions that we describe here. 

In Figure 13, we plot the upper-bound estimates of the He production cross sections from the data against
the multiplicity-weighted cross sections predicted by PHITS and LAQGSM for the two $^{14}$N beams and four 
$^{16}$O beams (left), and separately for the three $^{20}$Ne beams and one $^{24}$Mg beam (right). The 
experimental error bars are shown. For the $^{16}$O beams incident on H, C, Al, and Cu targets (measured 
cross sections below 1100 mb), the agreement with PHITS is good, while the measured cross sections for the 
Sn and Pb targets are systematically higher than predicted by PHITS by about one standard error. The comparisons 
shown here are to the upper-bound estimates from the data, and looking at only the $^{16}$O data, one might conclude 
that this is what causes almost all of the data points to fall above the 45$^\circ$ line. However, the situation is 
reversed for the $^{20}$Ne beams: almost all of the predicted cross sections are larger than the measured cross 
sections, for both models, and the LAQGSM predictions are generally closer to the data than are the PHITS predictions. 

The good agreement between the measured upper-bound values and PHITS predictions for $^{16}$O beams is remarkable, 
but given that a similar level of agreement is not seen with $^{20}$Ne and $^{24}$Mg beams, we conclude that it is
likely fortuitous. However, in general, it is worth noting that the data and the predictions of the two models are not 
grossly in disagreement. The dotted lines in Figures 13(a) and 13(b) show factor-of-two errors; in the left-hand plot, 
all points fall between the lines, and in the right-hand plot, five H-target data points fall outside the lines while 
all other points are within.

\begin{figure*}
\includegraphics[width=4.5in]{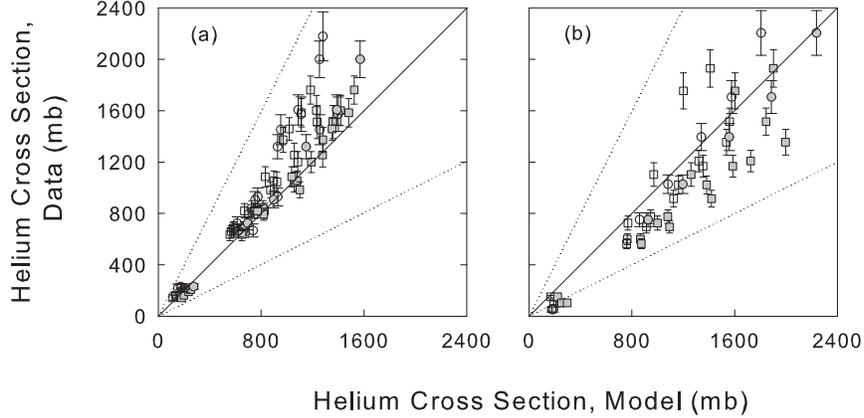}
\caption{\label{fig:fig13}Upper-bound estimates for helium production cross sections plotted vs. PHITS and LAQGSM
predictions, for C, Al, Sn, and Pb targets. Results for $^{16}$O beams are shown in Fig. 13(a) on the left, and for 
$^{20}$Ne and $^{24}$Mg beams in Fig. 13(b) on the right. For all beams, open symbols are used for LAQGSM and 
filled symbols for PHITS; in the left-hand plot, square symbols are for $^{16}$O beams, and in the right-hand plot, 
square symbols are for the $^{24}$Mg beam. The dotted lines correspond to factor-of-two differences between the 
data and the models.}
\end{figure*}

\subsection{Cumulative Relative Uncertainty}
Employing the bounding method for helium production cross sections, and taking the upper-bound estimates,
we have 408 fragment production cross sections available for model comparisons. A concise method for 
comparing data and model predictions has recently been described by Norman and Blattnig \cite{Norman2008}.
Here, we briefly recapitulate the method and apply it to the ten data sets presented above. Comparisons are 
performed for NUCFRG2, PHITS, and LAQGSM for fragment species from one charge unit below the primary down
to helium. The three quantities needed to implement the method are defined as follows.

\[D^+\equiv \frac{M\left ( x_i \right )-\left [ E\left ( x_i \right ) + \epsilon \left ( x_i \right )\right ]}{E\left ( x_i \right )+\epsilon \left ( x_i \right )}\]

\[D^-\equiv \frac{M\left ( x_i \right )-\left [ E\left ( x_i \right ) - \epsilon \left ( x_i \right )\right ]}{E\left ( x_i \right )-\epsilon \left ( x_i \right )}\]

\[U\left ( x_i \right )\equiv \textup{MAX}\left ( \left | D^+\left ( x_i \right ) \right |,\left | D^-\left ( x_i \right ) \right | \right )\]

The $D$ value for a given data point $i$ depends on the cross section predicted by the model, $M(x_i)$, the cross section
as measured in the experiment, $E(x_i)$, and the uncertainty associated with the measurement, $\epsilon(x_i)$. 
Values of $D$ were calculated for each of the 408 data points available. For each model, for helium, comparisons
to the upper-bound values described above yield the lowest average $U(x_i)$ values.

Figure 14 shows the cumulative probability distributions of the relative uncertainties as defined above, calculated
for NUCFRG2, PHITS, and LAQGSM. This graph should be read from left (0.0) to right (1.0).
All three models yield similar curves for about 25\% of the data, at which point the 
NUCFRG2 curve rises slightly above the other two. The PHITS and LAQGSM curves lie practically on top of one another,
below the NUCFRG2 curve, from about 0.25 to about 0.70. There, the LAQGSM curve begins to rise, crossing
the NUCFRG2 curve around 0.76. The LAQGSM curve remains above the NUCFRG2 curve, which remains slightly
above the PHITS curve. The ordering persists out to about 0.99, where all three curves converge at a $U(x_i)$ 
value of about 1.6. Based on these curves, PHITS can be said to have the best overall agreement with the data.

\begin{figure}
\includegraphics[width=3.35in]{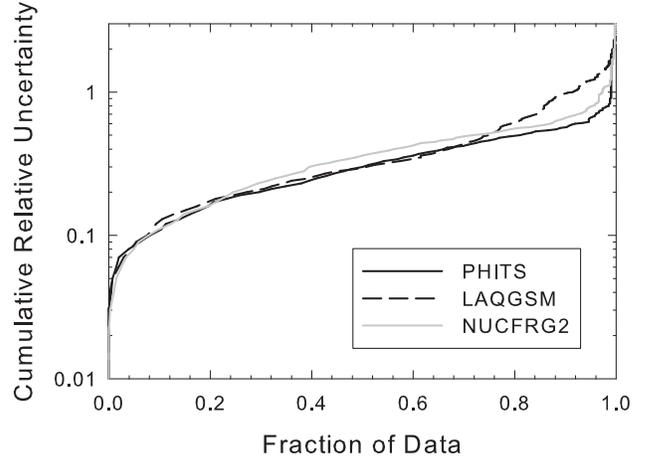}
\caption{\label{fig:fig14}Cumulative relative uncertainty as defined in the text, calculated for the NUCFRG2, PHITS,
and LAQGSM models.}
\end{figure}

Interpreting the curves in Figure 14 is somewhat subjective. On one hand, we can say that with any of the models
under consideration, about 70\% of the cross sections have $U(x_i)$ values less than 0.5. More critically, we can say 
that only about 30\% of the predicted cross sections are highly accurate ($U(x_i)$ values less than 0.25) and that 
values of $U(x_i)$ above 1.0 represent significant weaknesses of the codes in those cases where the experimental
errors are modest. (Large experimental errors dominate the $U(x_i)$ in a few cases here.) Of course the definition 
of what constitutes a ``good'' value of $U(x_i)$ is arbitrary and it could be argued that any values above, say, 0.5 
are problematic. Figure 14 shows that, at least for these beams, PHITS is slightly more accurate overall than the 
other two models tested for these beam ions and energies. 

\subsection{Model Systematics}
The cumulative relative uncertainty curves shown in Figure 14 give an overall picture of how well the models reproduce 
the data, but there are
details and systematic differences that cannot be conveyed in such a compact format. A highly detailed discussion 
of these comparisons is beyond the scope of this article, but we will describe some obvious features that
are noteworthy. We begin with LAQGSM, which, as shown in Figure 14, deviates from the other two models considered here
once we get to about 70\% of the data. A closer look at the source of the discrepancies reveals that they are largely
confined to two fragment categories, charge 3, and $\Delta Z$ = 1. To illustrate this, in Figure 15 we show the $U(x_i)$ 
values for LAQGSM predictions of N production cross sections vs. target mass for the four $^{16}$O beams. Clearly, 
LAQGSM does not agree well with the data for the lighter targets, and does somewhat better (but still not well) 
as target mass increases. In all cases, $U(x_i)$ is above 0.5, indicating a lack of agreement with the data. The 
situation is similar, albeit not quite as bad, for $\Delta Z$ = 1 with the other beam ions. With the exception of
the $^{14}$N data, these cases test the INC part of LAQGSM. For Li production, which in LAQGSM depends on the Fermi 
break-up model, half of the values of $U(x_i)$ are above 1.0, and several more are above 0.9. While the $U(x_i)$ values 
for Li production tend to be comparatively large for all models due to the large experimental uncertainty, the LAQGSM 
comparison yields particularly large values, because (unlike the other models) it predicts Li cross sections that are 
in all cases significantly larger than the cross sections estimated from the data. With the unavoidable large uncertainties 
assigned to these data points, the $D^-$ values (corresponding to the case where the estimated cross section is smallest) 
are quite large. 

With sixty data sets (ten beams and six target materials), the cross sections for Li and for $\Delta Z$ = 1 account for 
120 points of comparison out of 408. The lack of agreement for these cases largely explains the rise of the LAQGSM curve 
in Figure 14 when the data fraction reaches 0.7.

\begin{figure}
\includegraphics[width=3.35in]{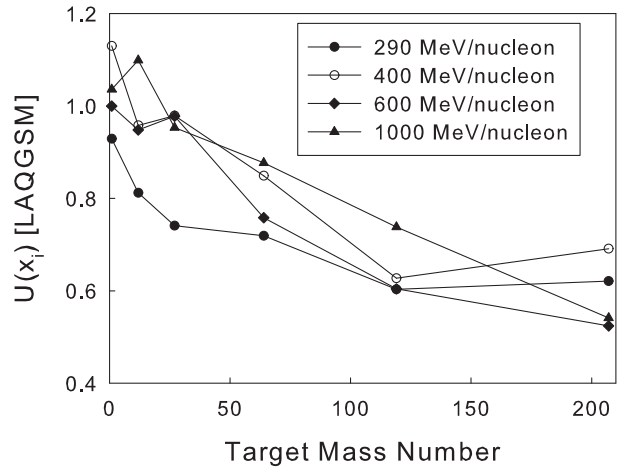}
\caption{\label{fig:fig15}Values of relative uncertainty $U(x_i)$ obtained with the LAQGSM model for the reaction
$^{16}$O + A $\rightarrow$ N + X for the four beams and six targets used in the measurements.}
\end{figure}

For NUCFRG2, about 30\% of the cross sections predicted by this model yield a $U(x_i)$ value above 0.5, as can be 
seen in Figure 14. The NUCFRG2 curve shows an inflection point at a data fraction of about 0.94, followed by a fairly
steep rise, largely driven by fourteen data points that have $U(x_i)$ values greater than 0.90. Ten of the fourteen
discrepant data points are for hydrogen targets; five of these ten are for helium production. Of the other four highly
discrepant points, three are for F production ($Z_{\text{frag}}$ = 9) in the $^{24}$Mg beam data. In all fourteen cases, 
NUCFRG2 predicts cross sections that are roughly factors of 2 larger than the measured cross sections.

For PHITS, only about 20\% of the cross sections have $U(x_i)$ values greater than 0.5, and only 5 of the 408 
cross sections yield values above 0.90. Most of the large $U(x_i)$ values are seen are for Li production, where the 
experimental uncertainty contributes most of the total. 

All models yield large values of $U(x_i)$ for the helium production cross section with $^{24}$Mg on H. The upper-bound
estimate is 52 $\pm$ 20 mb, and the three model predictions are remarkably consistent for this point. NUCFRG2 predicts
a cross section of 176 mb, LAQGSM 178 mb, and PHITS 189 mb. All are more than a factor of three larger than the estimate
from the data. This is the most extreme example of a broader trend for H targets, in which all three models predict 
substantially larger He production cross sections than are estimated from the data. 

The mean values of $U(x_i)$ for the 408 cross sections considered here are 0.35, 0.40, and 0.43 for PHITS, NUCFRG2, and 
LAQGSM, respectively. The order of the results (PHITS smallest, LAQGSM largest) is predictable given the curves in Figure 14. 
Other comparisons between PHITS, LAQGSM, and data have previously been reported, including measurements
of neutron spectra \cite{iwase2005}, proton-induced reactions\cite{titarenko2011}, fragmentation of $^{28}$Si \cite{sihver2008},
and fragmentation of $^{12}$C and $^{56}$Fe \cite{Mashnik2008b}. (Note that results reported as being for the MCNPX code use 
LAQGSM.) Overall, LAQGSM tends to be at least as accurate as PHITS, and in some cases it is considerably more accurate. 
The results of the analysis presented here pertain to light-ion beam species; further investigation using cross section
data for heavier beams is needed.

\subsection{Comparison to Webber et al. Data}
The Cumulative Relative Uncertainty method can also be used to compare a subset of the large-acceptance fragment 
production cross sections to those obtained by Webber et al. \cite{Webber1990b}. In the equations for $D^+$ and $D^-$ above, 
we take the $M(x_i)$ terms to be the cross sections reported in the earlier work, the $E(x_i)$ to be our cross sections as per 
Tables VI through IX, and the $\epsilon(x_i)$ to be the quadrature sums of the uncertainties reported by the two groups. 
There are 54 data points that can be reasonably compared. We find an average $U(x_i)$ of 0.154 and no values 
greater than 0.533. Although this level of agreement suggests lurking systematic errors beyond those claimed, it is still far
better than the agreement between the data and any of the models. 

Fragment production cross sections with isotopic resolution using oxygen beams at comparable energies have been reported 
by Leistenschneider et al. \cite{Leistenschneider2002} (at $\approx$ 600 MeV/nucleon) and Momota et al. \cite{momota2002}
(at 290 MeV/nucleon). However, comparisons to those experiments are not straightforward owing to large differences
in the detector configurations and the methods of extracting fragment production cross sections. The Leistenschneider et al.
data have been compared to a previous version of LAQGSM \cite{Mashnik2003}.

\section{Conclusions}
We have presented 60 charge-changing cross sections and over 500 fragment production cross sections for beams of
$^{14}$N, $^{16}$O, $^{20}$Ne, and $^{24}$Mg at energies ranging from 290 to 1000 MeV/nucleon. Comparisons to the
PHITS, NUCFRG2, and LAQGSM models have been made. PHITS and NUCFRG2 can be directly compared to the measured
charge-changing cross sections, and show reasonable agreement with the data. A simple geometric model with two free 
parameters (the nucleon radius and the overlap term) can be tuned to fit the bulk of the charge-changing cross section 
data slightly better than the more sophisticated models. 

Large-acceptance charge spectra are used to extract fragment production cross sections for charges above about
half the beam charge. These cross sections are corrected for interactions in the detectors and intervening materials,
and for secondary interactions in the target; they do not require acceptance corrections. Small-acceptance charge
spectra are used to measure cross sections for lighter fragments, and these do require acceptance corrections
that monotonically increase with the charge change ($\Delta Z$).

The small-acceptance charge spectra are found to invariably contain peaks in the region of $Z \approx 3.5$. We have 
reported these peaks previously; they appear to be events in which three helium fragments are detected simultaneously.
Here, by comparing cross sections for this peak at different beam energies and acceptances, we find that the data are 
best explained by production of $^8$Be, which instantly decays into two $^4$He, in association with a third, 
independently-produced He fragment.

We presented cross sections for three categories of events that are either partially or entirely due to helium fragments, 
which are copiously produced in these reactions. When the measured cross sections are combined in a multiplicity-weighted 
fashion to yield upper- and lower-bound estimates for total helium production cross sections, agreement with PHITS and LAQGSM
predictions is good. The data and both models agree within a factor of 2 in almost all cases. Good agreement is seen 
when PHITS is compared to the N and O beam data, and when LAQGSM is compared to the Ne and Mg beam data. 

Four hundred eight fragment production cross sections presented here were used for comparisons to the three models 
using the Cumulative Relative Uncertainty method. All three models yield similar results, as seen in Figure 14; the models
can all be said to be reasonably accurate ($U(x_i) < 0.5$) for 75\% of the fragment production cross sections. Overall,
PHITS gives the best overall agreement, followed by NUCFRG2 and LAQGSM. To the limited extent that comparable
data are available, the present data are in reasonable agreement with earlier measurements. 

\begin{acknowledgments}
We thank the accelerator operators and support teams at the HIMAC facility at NIRS, and at the NSRL facility at BNL. 
Their efforts make this work possible. We are particularly grateful to the HIMAC PAC for their extreme generosity 
in granting us hundreds of hours of beam time at no charge. 

This work was supported at Southwest Research Institute by NASA Grant Number NNX09AE18A. At LBNL, this work was
supported by the Space Radiation Health Program of the National Aeronautics and Space Administration under NASA Grant 
Numbers L14230C and H31909D, through the U.S. Department of Energy under Contract No. DE-AC03076SF00098. At HIMAC, 
this work was supported in part by the Research Project with Heavy Ions at NIRS-HIMAC, Project No. P037. The part of
this work performed at LANL was carried out under the auspices of the National Nuclear Security Administration of the 
U.S. Department of Energy at Los Alamos National Laboratory under Contract No. DE-AC52-06NA25396 with funding from 
the Defense Threat Reduction Agency (DTRA).

\end{acknowledgments}

\bibliography{refs}

%\printfigures

\end{document}